\documentclass[fleqn,10pt]{wlpeerj}
\usepackage{tabularx}
\usepackage{longtable}
\usepackage{url}
\usepackage{subcaption}
\usepackage{graphicx}
\usepackage{soul}
\usepackage[many]{tcolorbox}
\usepackage[hidelinks]{hyperref}
\usepackage{tikz}
\usetikzlibrary{positioning}
\usepackage[toc,page]{appendix}

\title{Evaluating the method reproducibility of deep learning models in the biodiversity domain}
\author[1,5]{Waqas Ahmed}
\author[1,3,4,5]{Vamsi Krishna Kommineni}
\author[1,2,4]{Birgitta K\"onig-Ries}
\author[1]{Jitendra Gaikwad}
\author[1]{Luiz Gadelha}
\author[1,2]{Sheeba Samuel}
\affil[1]{Heinz Nixdorf Chair for Distributed Information Systems, Friedrich Schiller University Jena, Ernst-Abbe-Platz 2, Jena,07743, Thuringia, Germany}
\affil[2]{Michael Stifel Center Jena, Friedrich Schiller University Jena, Leutragraben 1, Jena, 07743, Thuringia, Germany}

\affil[3]{Department of Functional Biogeography, Max Planck Institute for Biogeochemistry, Hans-Knoell-Str. 10, Jena, 07745, Thuringia, Germany}

\affil[4]{German Centre for Integrative Biodiversity Research (iDiv) Halle-Jena-Leipzig, Puschstraße 4, Leipzig, 04103, Saxony, Germany}
\corrauthor[1]{Waqas Ahmed}{ahmed.waqas@uni-jena.de}

\begin{abstract}

Artificial Intelligence (AI) is revolutionizing biodiversity research by enabling advanced data analysis, species identification, and habitats monitoring, thereby enhancing conservation efforts. 
Ensuring reproducibility in AI-driven biodiversity research is crucial for fostering transparency, verifying results, and promoting the credibility of ecological findings.
This study investigates the reproducibility of deep learning (DL) methods within the biodiversity domain.  
We design a methodology for evaluating the reproducibility of biodiversity-related publications that employ DL techniques across three stages.
We define ten variables essential for method reproducibility, divided into four categories: resource requirements, methodological information, uncontrolled randomness, and statistical considerations.
These categories subsequently serve as the basis for defining different levels of reproducibility.
We manually extract the availability of these variables from a curated dataset comprising 61 publications identified using the keywords provided by biodiversity experts.
Our study shows that the dataset is shared in 47\% of the publications; however, a significant number of the publications lack comprehensive information on deep learning methods, including details regarding randomness.
\end{abstract}

\begin{document}

\flushbottom
\maketitle
\thispagestyle{empty}

\section*{Introduction}
\label{sec:Introduction}

In recent years, deep learning methods have been increasingly applied to understand complex ecological systems, particularly in the field of biodiversity. These methods have the potential to process and analyze large amounts of biological data rapidly, leading to significant insights. For instance, \cite{AUGUST2020100116} demonstrated how AI image classifiers can create new biodiversity datasets from social media imagery, highlighting the spatial and taxonomic biases that can influence ecological inference. Similarly, DL models have been utilized to analyze camera trap images for wildlife monitoring, enabling researchers to identify species and infer ecological patterns and processes with high accuracy in \cite{tabak2019machine}. However, there is a growing concern about the reproducibility, transparency, and trustworthiness of research findings produced using deep learning methods in this domain \citep{feng2019checklist, gpai2022biodiversity}.

Reproducibility is essential in scientific research as it allows researchers to validate and advance methods and results, ensuring the reliability of scientific claims \citep{samuel2021understanding}. \cite{goodman2016what} define research reproducibility in three aspects: methods reproducibility, results reproducibility, and inferential reproducibility. In this paper, we focus on methods reproducibility, which involves the ability to exactly replicate the experimental and computational processes using the same data and tools to achieve the same outcomes. This is especially important in biodiversity conservation, where decisions can directly impact ecological health.
For example, \cite{norouzzadeh2018automatically} demonstrated the application of deep learning for automatically identifying and counting wildlife in camera trap images, which is vital for monitoring species populations and informing conservation strategies. However, failure to reproduce these methods accurately could lead to erroneous population estimates, thereby compromising conservation efforts and resource allocation. Similarly, in the study by \cite{rovero2013camera}, the deployment of camera traps for wildlife monitoring highlighted the need for reproducible methods to ensure consistent data collection and analysis across different geographical locations. Inconsistent methods could result in data that are not comparable, leading to flawed conclusions and ineffective conservation measures.
Moreover, the inability to reproduce results due to methodological inconsistencies can prevent the detection of errors that might hide underlying biases or issues in the data. For instance, \cite{christin2019applications} discuss how the reproducibility of deep learning models in ecological research is challenged by the complexity and heterogeneity of biodiversity data, which includes interactions between variables, missing values, and non-linearity. Ensuring reproducibility helps to uncover these complexities and improve the reliability of ecological models used for decision-making.


A deep learning pipeline is a structured sequence of processes in training and deploying a DL model \citep{el2020deep}.
The pipeline typically begins with data collection and preprocessing, involving tasks such as data cleaning, normalization, and transformation. Following data preprocessing, the next stage consists of designing and selecting an appropriate deep learning architecture, considering factors like model complexity and the nature of the problem. Subsequently, model training takes place, where the chosen architecture is trained on the preprocessed data using optimization algorithms and specific hyperparameter configurations. After training, the model is evaluated and fine-tuned using various performance metrics to ensure its effectiveness in solving the targeted problem. The best performing model is run on the test
data. To ensure an unbiased evaluation of the model's predictive performance. Finally, the trained model is deployed for real-world applications or further refinement.

To ensure the reproducibility of the deep learning pipeline, comprehensive documentation is crucial at each stage. This includes detailed records of the data collection steps (including providing persistent identifier for each data point), data preprocessing steps, such as the specific data transformation techniques applied and any data augmentation strategies employed. Additionally, it is vital to document the specifics of the chosen deep learning architecture, including the exact configurations and versions of the neural network layers utilized. Detailed notes on the hyperparameter values selected during the model training phase are essential, as well as the training convergence criteria and the optimization algorithms employed. Proper documentation of the evaluation metrics used and the testing dataset ensures the reproducibility of the model's performance assessment. Finally, maintaining a record of the software libraries, hardware, frameworks, and versions utilized throughout the pipeline aids in replicating the experimental setup accurately.

In this paper, we aim to shed light on the current state of reproducibility of deep learning methods applied to biodiversity research. 
We conducted a systematic literature review to identify publications that use deep learning techniques in biodiversity research using keywords provided by biodiversity experts \citep{abdelmageed2022biodivnere}. We define various reproducibility-related variables inspired by the current literature. We then curated a dataset of 61 publications and manually extracted reproducibility-related variables, such as the availability of datasets and code. We also analyzed advanced reproducibility variables, such as the specific deep learning methods employed, models, and hyperparameters.
Our findings suggest that the reproducibility of deep learning methods in biodiversity research is generally low. However, there is a growing trend towards improving reproducibility, with more and more publications making their datasets and code available. Our study will contribute to the ongoing discourse on the reproducibility of deep learning methods in biodiversity research and help to improve the credibility and impact of these methods in this vital field.

In the following sections, we provide a detailed description of our study. 
We start with an overview of the state-of-the-art (``Related Work'').
We provide the methodology of our study (``Methododology'')
We describe our results and discuss the implications of our work (``Discussion'').
Finally, we summarize the key aspects of our study and provide future directions of our research (``Conclusion'').

\section*{Related Work}
\label{sec:relatedwork}
Method reproducibility, as mentioned in \cite{goodman2016what}, is an important aspect of progress in any scientific field. \cite{raff2019step} provides some insight into reproducibility in ML. The author studied the reproducibility of 255 ML papers published from 1984 to 2017 and tried to correlate the reproducibility of the papers with their characteristics. According to his observation, the main obstacle to reproducibility of results is insufficient explanation of the method and implementation details when the source code is not provided. However, some of the defined attributes were very subjective, such as the algorithmic difficulty of the work or the readability of the work. While \cite{raff2019step} provides insight into ML reproducibility in general, there is also research specifically related to the field of biodiversity \citep{feng2019checklist,schnitzer2016would}. They pointed out that reproducibility is problematic because erroneous data is widespread and there is a lack of empirical studies to verify previous research findings. Several authors have pointed out the need for better data management and reporting practices to ensure the reproducibility of research results \citep{stark2018before,waide2017demystifying,samuel2020machine}. In \cite{waide2017demystifying}, the authors discuss the challenges associated with managing different data sets in ecological research, while \cite{leonelli2018rethinking} further argues that reproducibility should be a criterion for assessing research quality. However, setting standards for assessing reproducibility itself is an important issue for research. \cite{gundersen2022machine} in his latest study defined 22 variables that could be categorized according to different degrees of reproducibility. Three categories were mentioned: Reproducible Data, Reproducible Experiment and Reproducible Method. \cite{gundersen2022machine} used these variables to evaluate reproducibility support on 13 different open source machine learning platforms. Similarly, \cite{heil2021reproducibility} proposed 3 different standards for reproducibility in machine learning that have applications in the life sciences. Bronze, silver or gold standards are defined according to the availability of data, models and code. \cite{tatman2018practical} went a step further and analyzed ML papers from ICML and NeurIPS conferences and distinguished levels of reproducibility based on the resources provided with the paper. To achieve the desired result, they recommended some practical steps such as providing code and data in an executable environment where all libraries and dependencies are linked. This allows code or experiments to run smoothly on a new machine. This recommendation is in line with the gold standard provided in \citep{heil2021reproducibility} that authors should produce reproducible results with the execution of a single command. In addition to the general recommendations, \cite{pineau2021improving} proposed standard practices and activities to improve reproducibility in the AI community. Some of these are the reproducibility programme at the NeurIPS conference and the ML reproducibility checklist, which provides guidelines to authors before submitting to conferences. Inspired by \cite{pineau2021improving}, some of the major AI conferences (AAAI and IJCAI) have introduced similar reproducibility checklists. We have considered these guidelines and also previous work in \cite{gundersen2022machine} to develop 10 reproducibility variables. However, the work in \cite{gundersen2022machine} is designed to assess the reproducibility of different ML platforms, whereas we aim to assess the reproducibility of biodiversity research. Our variables also include the aspect of uncontrolled randomness and the statistical information necessary for reproducibility.

\section*{Methodology} 
\label{sec:methodology}

In this section, we will discuss the steps of our work analyzing the reproducibility of research papers in the biodiversity domain.

\subsection*{Identification:} To assess the reproducibility of methods in biodiversity research, we first tried to obtain unbiased and relevant publications. To do this, we adopted keywords from \cite{abdelmageed2022biodivnere}, which were provided by biodiversity experts and were also used to develop a corpus in the field of biodiversity for large language models \citep{abdelmageed2022biodivnere}. These keywords are: "biodivers*" OR "genetic diversity" OR "*omic diversity" OR "phylogenetic diversity" OR "population diversity" OR "species diversity" OR "ecosystem diversity" OR "functional diversity" OR "microbial diversity" OR "soil diversity" AND "Deep Learning".

With this search query, Google Scholar suggested more than 8000 articles in a period from 2015 to 2021. However, we only selected the first 100 results for our analysis, as manual processing of such a large number of articles was not possible in a limited time frame and we believe that the characteristics of these papers are statistically not different from those of the complete set.  We acknowledge that the first 100 articles selected for our analysis might not represent an entirely unbiased sample of the search results. However, our selection covers a broad range of publishers and publication years, thus capturing a diverse cross-section of the available literature. Additionally, we manually reviewed each selected publication to extract relevant information on ten defined variables, dedicating approximately 40 minutes per paper. This thorough manual curation aimed to enhance the reliability of our dataset. This manually curated dataset will serve as a benchmark for developing an automated system to extract variable information, enabling future analyses to encompass a larger number of articles while maintaining methodological rigor and reproducibility.


\subsection*{Screening:} 
\begin{figure}
    \centering
    \includegraphics[width=0.75\linewidth]{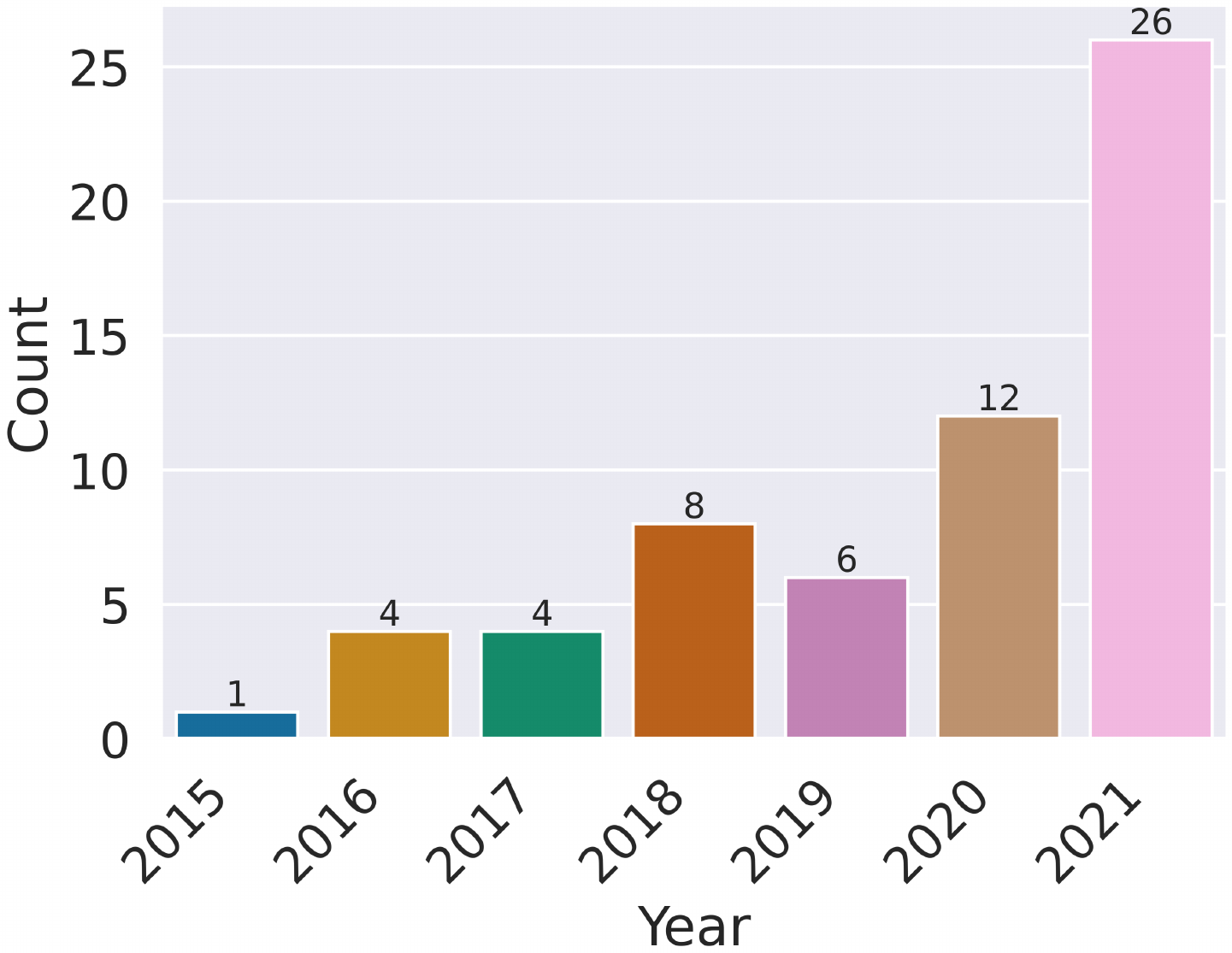}
    \caption{\centering
Number of publications considered for collecting the binary responses as per the reproducibility criteria based on year}
    \label{fig:year-info}
\end{figure}

\begin{figure}
    \centering
    \includegraphics[width=\linewidth]{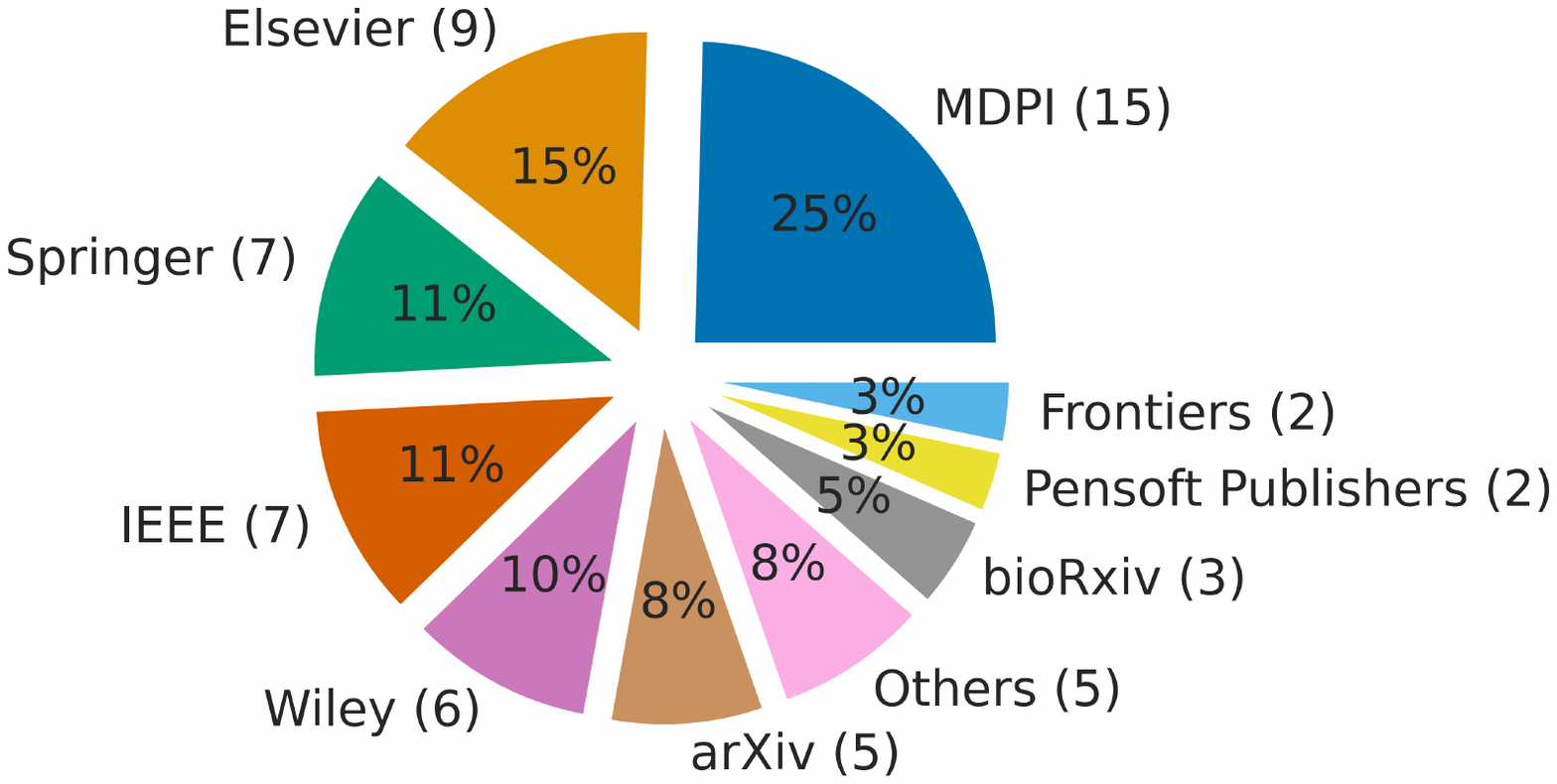}
    \caption{\centering
Publisher information for the 61 publications selected for collecting the binary responses as per the reproducibility criteria}
    \label{fig:pub-info}
\end{figure}

Before analyzing the individual articles for reproducibility, we found that some articles could not be considered further for the following reasons: 1) duplication, 2) only an abstract was provided or the full article was not accessible, and 3) some articles were not empirical studies and therefore did not include experiments. After considering all these limitations, the number of articles was reduced to 61. Figure~\ref{fig:year-info} shows the distribution of articles over the period from 2015 to 2021, with most articles being from  2020 and 2021. Figure~\ref{fig:pub-info} shows the publisher information for the considered articles, which are distributed quite randomly across more than ten publishers.

\subsection*{Selection of reproducibility variables:} There is no standard for assessing the reproducibility of DL biodiversity methods. A common practice for analyzing the reproducibility of research articles is to rerun the experiments using the same methodology as the original author. However, this requires a lot of time and computing resources. Also, sometimes it is not possible to work with the same resources such as hardware and software. Inspired by \citep{gundersen2022machine}, we have formulated a set of 10 indicators that can be used as proxies for the probability that a result is reproducible. Instead of repeating each experiment, we look for these variables or factors that are considered important determinants of the reproducibility. Using the literature and the reproducibility checklists at conferences such as NeurIPS and AAAI \citep{pineau2021improving}, we can divide these variables into four categories. Resource Information (ReI) details the availability of datasets, source code, open-source tools, and proprietary model details crucial for reproduction efforts. Methodological Information (MI) captures the specifics of software and hardware used, as well as a high-level overview of the deep learning methods employed. Randomness Information (RaI) addresses aspects of unpredictability in computational processes, ensuring they are documented for consistency. Statistical Information (SI) focuses on the rigor of result analysis, advocating for multiple metrics and averaging techniques for reliable evaluation. The comprehensive details that define these aspects of reproducibility are systematically itemized in Table~\ref{Tab:Def}
 
\subsection*{Reproducibility check:} 
\begin{figure}[h!]
  \centering
  \begin{tikzpicture}[scale=2, nodes={draw, rectangle, minimum size=1cm}]
    \node[fill=blue!30, label=right: {ReI}] (level1) at (2,4) {Level 1};

   \node[fill=red!30, label=right: {ReI+MI}] (level2) at (2,3) {Level 2};

    \node[fill=green!30, label=left: {ReI+MI+SI}] (level3) at (1,2) {Level 3};

    \node[fill=yellow!30, label=right: {ReI+MI+RaI}] (level4) at (3,2) {Level 4};

    \node[fill=orange!30, label=right: {ReI+MI+SI+RaI}] (level5) at (2,1) {Level 5};

    \draw (level1) -- (level2)
          (level2) -- (level3) -- (level5)
          (level2) -- (level4) -- (level5);
  \end{tikzpicture}
  \caption{Number of levels along with the categories covered for respective levels}
  \label{fig:lattice}
\end{figure}
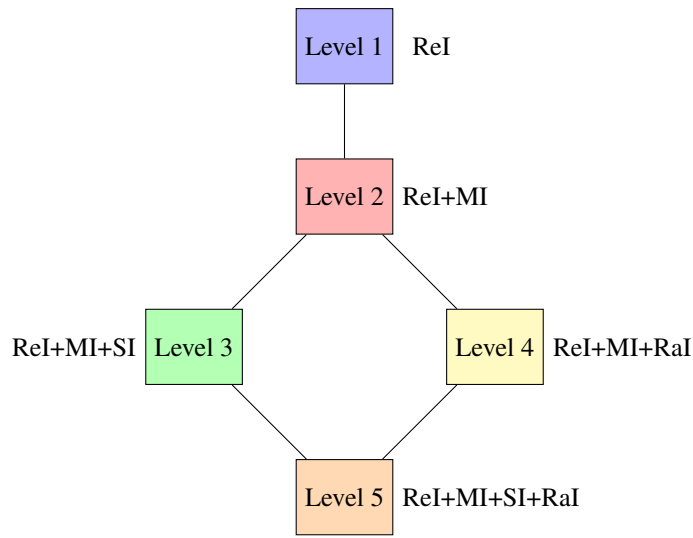

For all selected papers, we first manually checked the variables for each paper. To reduce the degree of subjectivity, we (two of the authors) began by independently recording the binary responses (availability or non-availability). In the initial phase of our assessment, we encountered notable discrepancies due to ambiguous interpretations of the definitions of each variable, leading to inconsistent binary responses. This was quantitatively evidenced by an average Cohen's Kappa value \citep{cohen1960coefficient} of 0.54, indicating a moderate level of agreement and highlighting the initial inconsistencies between annotators. To address these issues and improve the reliability of our analysis, we undertook a review and clarification of the variable definitions. In the end, we obtained the same binary responses for each paper, as shown in Table~\ref{Tab:Bin-res}. These responses formed the basis for analysing reproducibility. We verified whether the paper covered the functionality of each previously defined variable. After completing the binary responses for each paper, we quantified the reproducibility in 5 levels. The idea behind the different levels is to give independent researchers an insight into the chances of obtaining accurate results independently. The higher the level of reproducibility, the more variables are covered and the greater the chances of reproducing the result. These levels are set considering the time and computational effort required for reproducibility when certain variables are not present. However, the criteria for the basic level (Level 1) must be met to be reproducible according to the definition we refer to \cite{goodman2016what}. Figure~\ref{fig:lattice} shows the different levels of reproducibility together with the categories covered by each level. Level 1 should at least cover all variables defined in ReI. Level 2 should cover all variables defined in ReI and MI. Level 3 comprises the variables ReI, MI and SI, while Level 4 comprises the variables ReI , MI and RaI and these two levels do not build on each other. Finally, the highest level of reproducibility combines all categories at Level 5. 

\fontsize{10}{12}\selectfont
\renewcommand{\arraystretch}{1.5}
\begin{longtable}{p{2.2cm}p{2cm}p{1.5cm}p{6.5cm}}
\caption{Definition of the various reproducibility variables used in this research paper}
\label{Tab:Def} \\

\toprule
Category & Variable & Variable short name & Variable Description \\
\midrule
\endfirsthead

\multicolumn{4}{c}{{\tablename\ \thetable{} -- Continued from previous page}} \\
\toprule
Category & Variable & Variable short name & Variable Description \\
\midrule
\endhead

\bottomrule
\multicolumn{4}{r}{{Continued on next page}} \\
\endfoot

\bottomrule
\endlastfoot

Resource Information (ReI) & Dataset & V1 & Availability of dataset with persistent identifiers.  \\
 & Source Code & V2 & Availability of source code to recreate the experiment (e.g. GitHub, GitLab, Zenodo). \\
  & Open source frameworks or environment & V3 & Availability of open source software and hardware tools required to reproduce the work (e.g., docker container, virtual environments). \\
   & Model architecture & V4 & Availability of a deep learning model architecture or the accessibility of its internal working.   \\
Methodological Information (MI)& Software and Hardware Specification & V5 & Availability of information related to the type of hardware used (e.g., GPU), its specifications, and the version of software and libraries used. \\
   & Methods & V6 & Availability of high-level information about the deep learning pipeline including the pre-processing and post-processing steps. The other parameters of the pipeline are assessed as separate variables.\\
   & Hyper-parameters & V7 & Availability of hyperparameters used to train the model (e.g., number of epochs, learning rate, optimizer, etc.). \\
Randomness Information (RaI) & Randomness & R & Availability of information about weight initialization, data shuffling, data augmentation, data train-test split and cuDNN GPU library. \\
Statistical Information (SI) & Averaging result & S1 & Availability of multiple model training and averaging them out rather than selecting only the highest value. \\
 & Evaluation metrics & S2 & Availability of more than one evaluated metric (e.g., $R^2$ score along with Root Mean Square Error (RMSE) or Mean Absolute Error (MAE) or Loss of the model). \\

\bottomrule
\end{longtable}

\section*{Results} 

As discussed in Section~\ref{sec:methodology}, we divided the ten reproducibility variables we had defined into four categories 1) Resource Information 2) Methodological Information 3) Randomness Information, and 4) Statistical Information. These four categories were further categorized into five different levels of reproducibility, allowing us to assess the reproducibility level of the publications more comprehensively.

\label{sec:results}

\begin{figure}
    \centering
    \begin{subfigure}{0.48\textwidth}
        \centering
        \includegraphics[width=\linewidth]{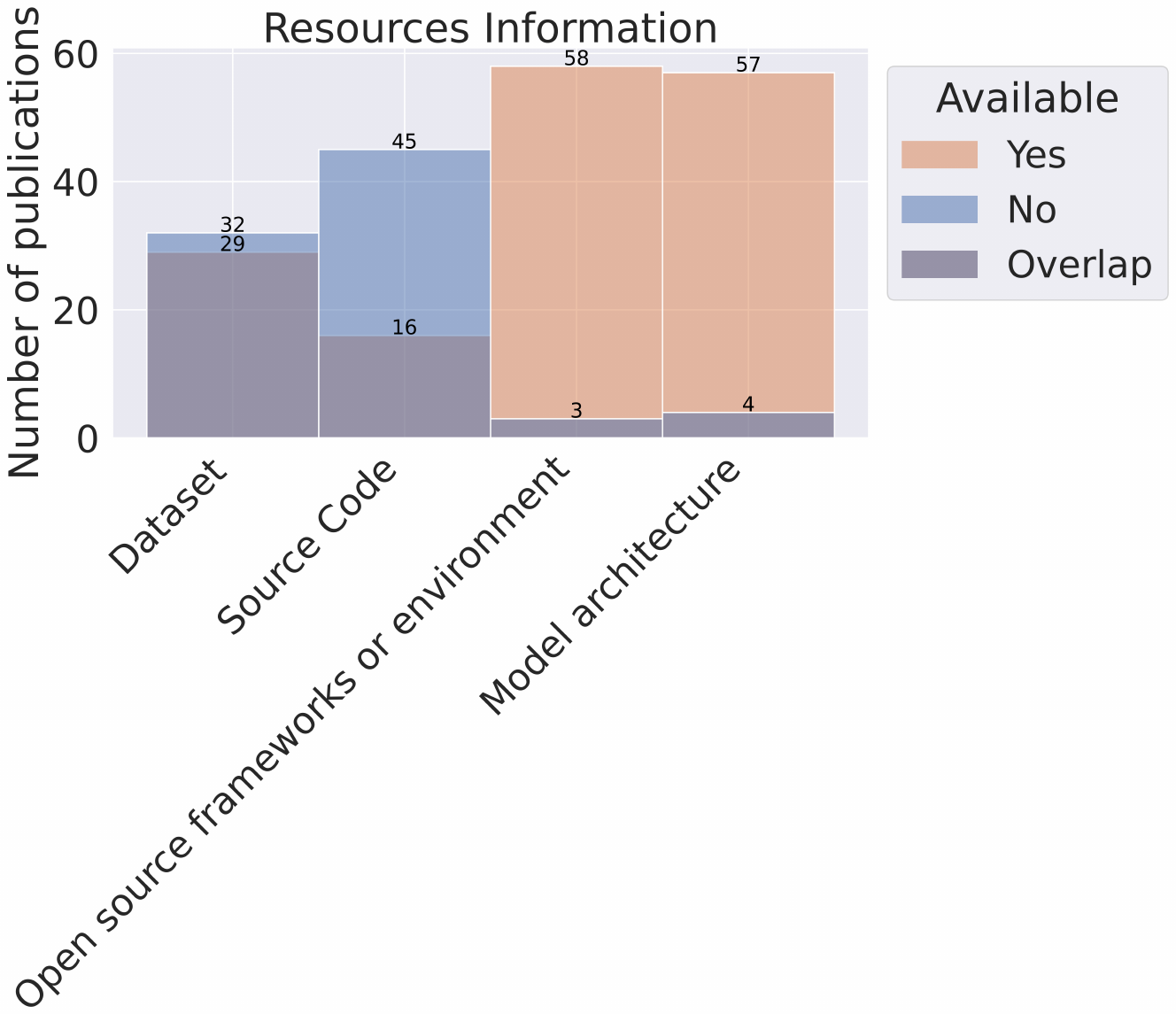}
        \caption{\centering Binary responses of variables for the category 'Resources'}
        \label{fig:subplot1}
    \end{subfigure}
    \hfill
    \begin{subfigure}{0.48\textwidth}l 
        \centering
        \includegraphics[width=\linewidth]{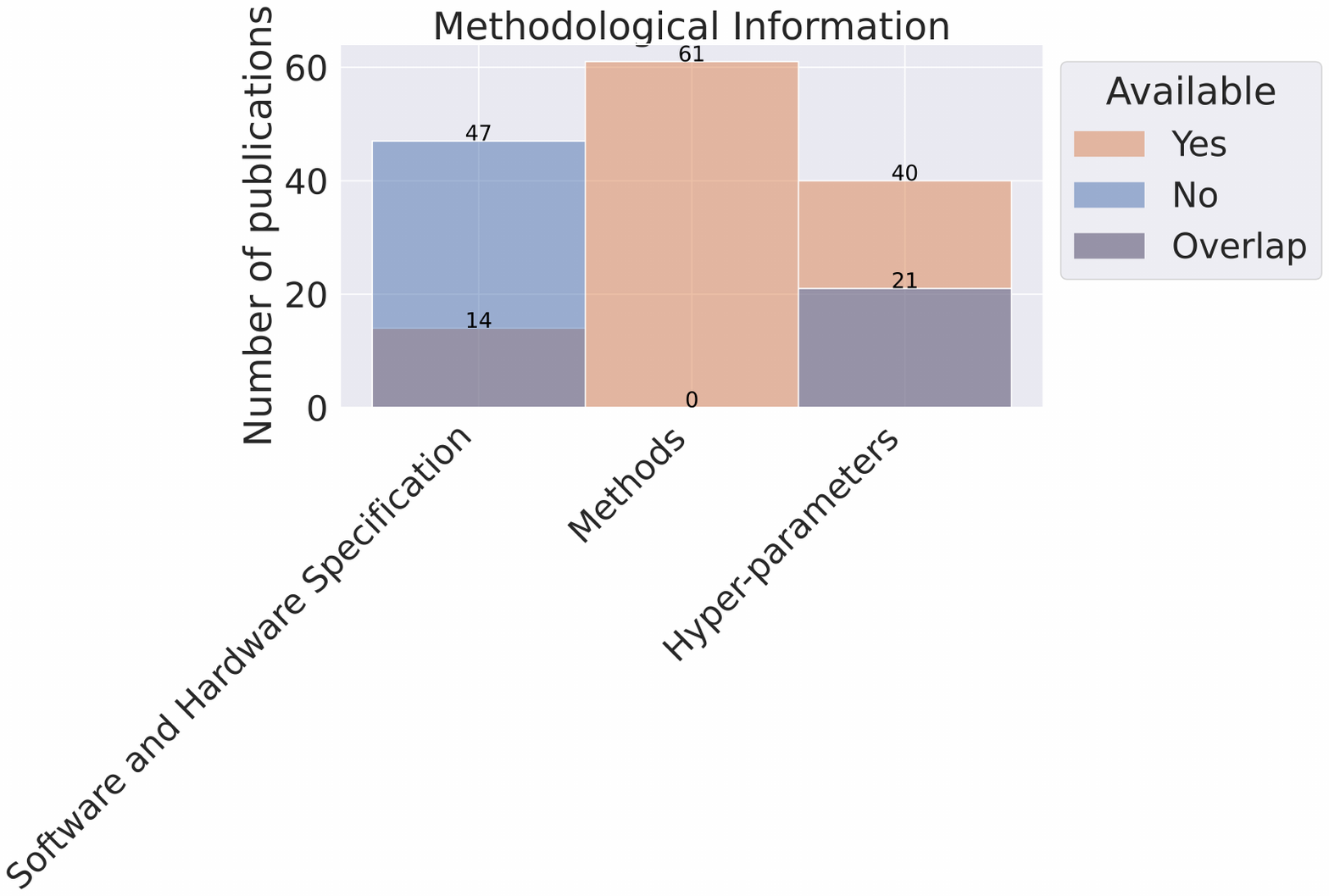}
        \caption{\centering Binary responses of variables for the category 'Methodological information'}
        \label{fig:subplot2}
    \end{subfigure}
    \centering 
    \begin{subfigure}{0.42\textwidth}
        \centering
        \includegraphics[width=\linewidth]{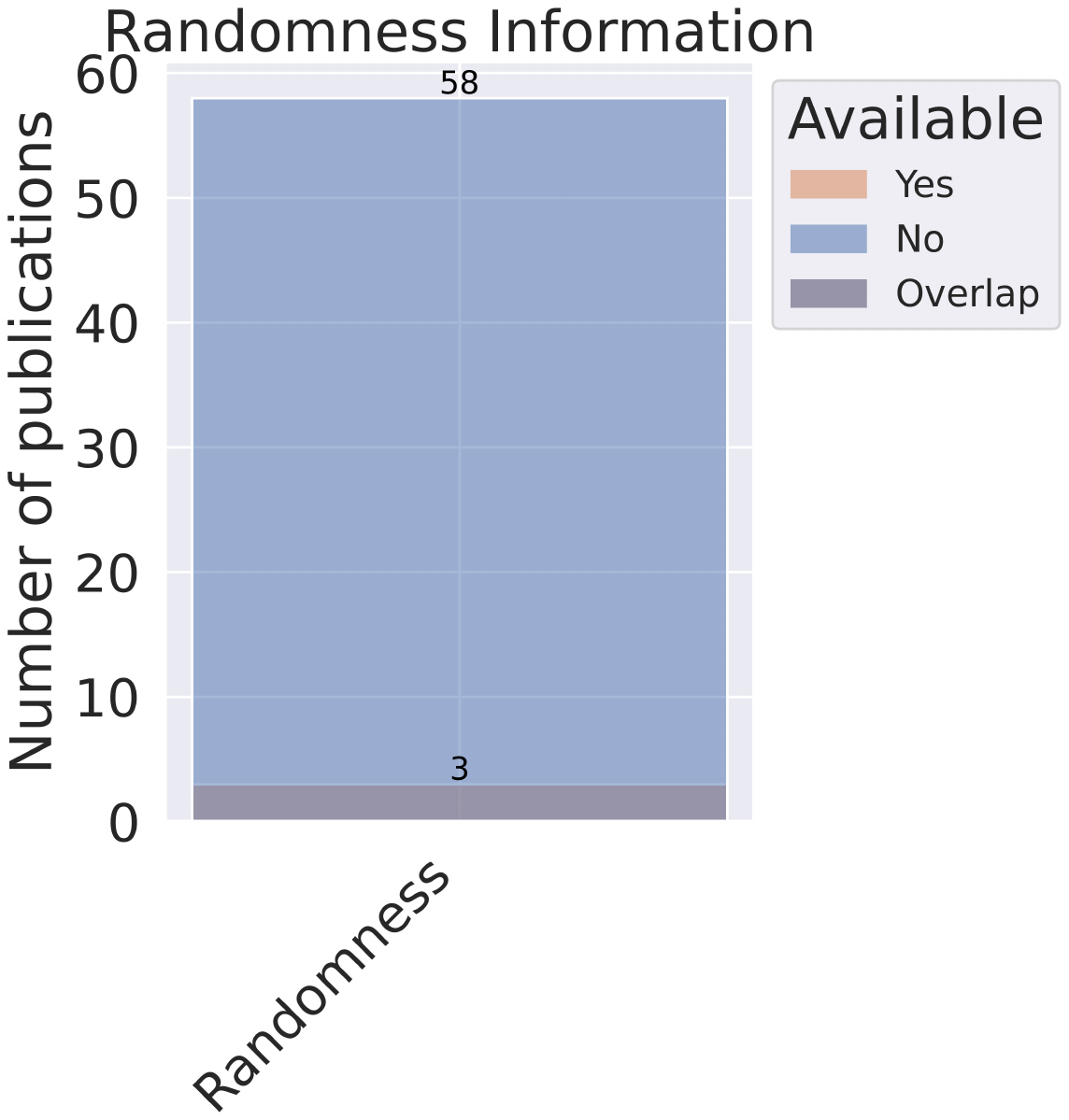}
        \caption{\centering Binary responses of variables for the category 'Randomness'}
        \label{fig:subplot3}
    \end{subfigure}
    \hfill
    \begin{subfigure}{0.48\textwidth}
        \centering
        \includegraphics[width=\linewidth]{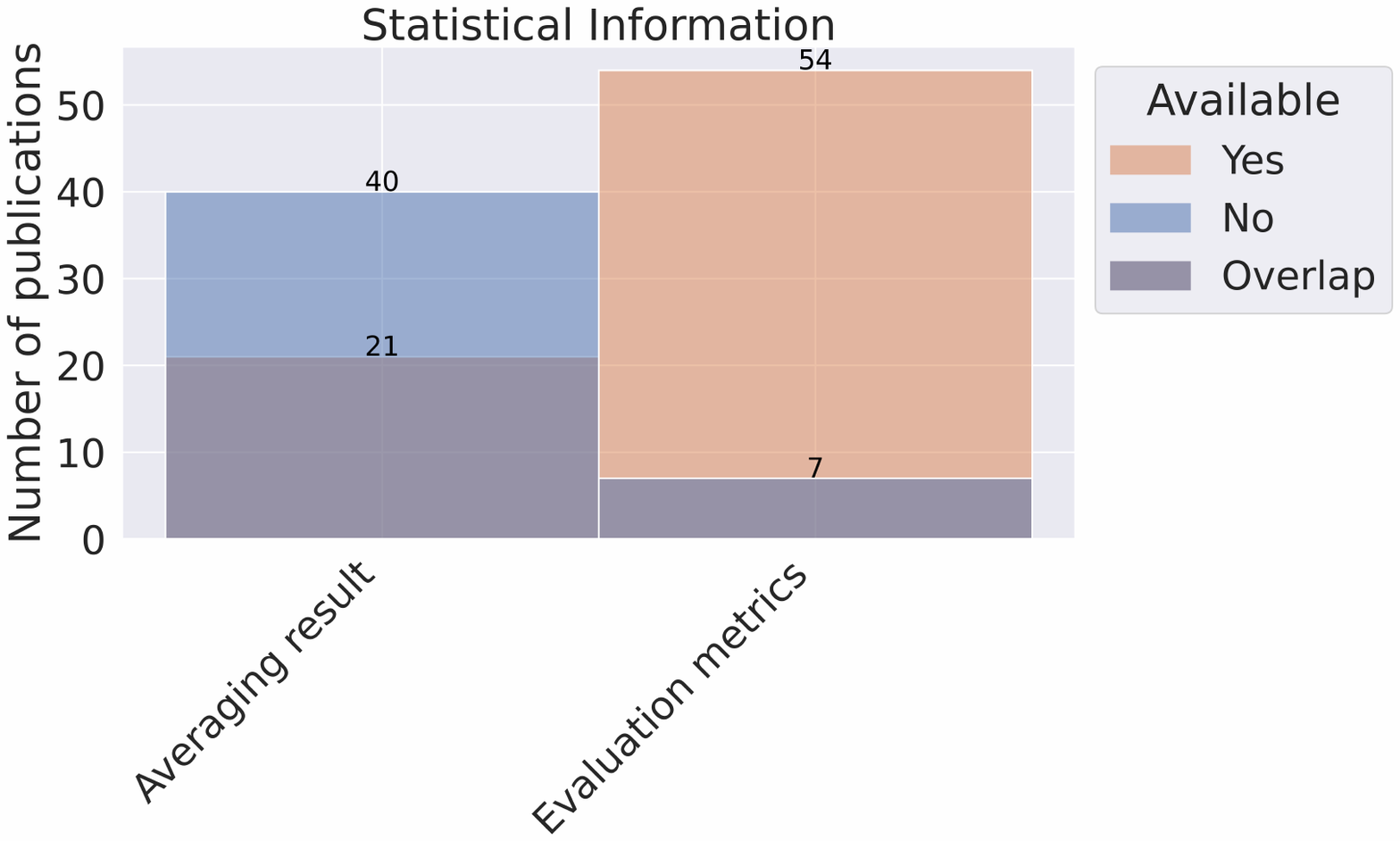}
        \caption{\centering Binary responses of variables for the category 'Statistical consideration'}
        \label{fig:subplot4}
    \end{subfigure}
    \caption{\centering
Binary responses of the considered reproducible variables in four categories 1) Resources 2) Methodological information 3) Randomness 4) Statistical consideration for selected research publication}
    \label{fig:binary-responses}
\end{figure}

\subsection*{Variable level information}
Figure~\ref{fig:binary-responses} contains four plots, one for each category of information. Within each category, specific reproducibility variables are depicted with binary counts, illustrated by two overlapping bars: one denoted by 'Yes' in a brick orange color and the other by 'No' in blue, along with the corresponding number of publications. In total, we made 610 individual judgements for the ten variables, 353 of these were positive i.e the respective information was provided, while this was not the case in 257 cases resulting in negative judgements. Detailed descriptions of all responses can be found at both the category and individual variable levels.

In the Resources Information category, 29 publications provided a dataset, 16 indicated the code repositories, 58 used open-source frameworks or environments, and 57 mentioned the model architectures (Figure~\ref{fig:subplot1}).

For the Methodological Information category, around a quarter of the publications included details about the hardware and the software (libraries) they used, all the publications explained methods that were used to build a machine learning pipeline and 40 publications provided the basic hyperparameters (Figure~\ref{fig:subplot2}).

The number of publications that used a random seed in all possible ways (weight
initialization, data shuffling, data augmentation, data train-test split and cuDNN
GPU library) in their code is 3 (Figure~\ref{fig:subplot3}). Regarding statistical considerations, 21 publications provided the average result of multiple model trainings, 54 publications evaluated their models with more than one evaluation metric. (Figure~\ref{fig:subplot4}).

\begin{figure}
    \centering
    \includegraphics[width=\linewidth]{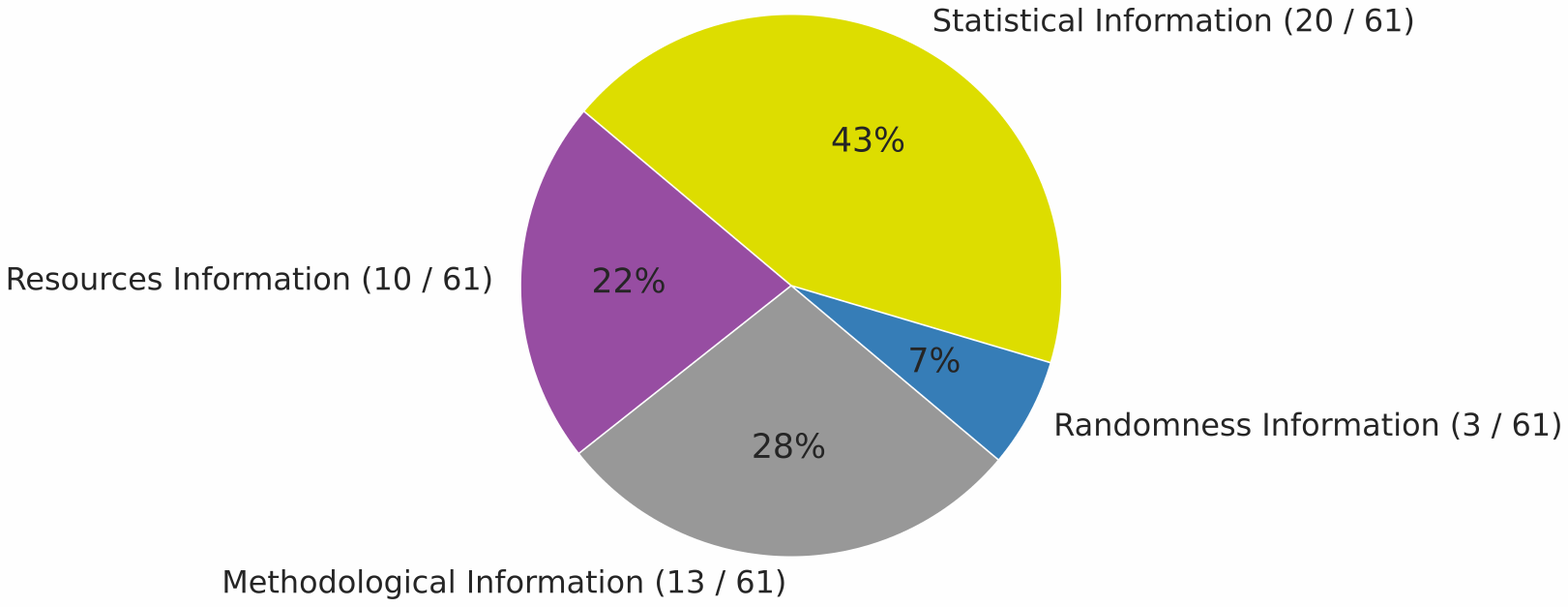}
    \caption{\centering
Distribution of selected research publications along four categories}
    \label{fig:cat-info}
\end{figure}

\subsection*{Categorical level information}
We have harmonized the binary responses of reproducibility variables for each categorical level in such a way that all the individual reproducibility variables in a category must be available (Yes) to mark that the specific categorical information is available (Yes). If one of the individual reproducibility variables is unavailable (No), the specific categorical information is unavailable (No). Mathematically, \(V1 \& V2 \& V3 \& V4 == Yes\) is used to denote the availability of resource information (Yes). 

Figure~\ref{fig:cat-info} depicts the distribution of publications that satisfy the defined categories. The number of publications that satisfy the Resources, Methodological, Randomness, and Statistical information are 10, 13, 3, and 20, respectively.

\begin{figure}
    \centering
    \includegraphics[width=\linewidth]{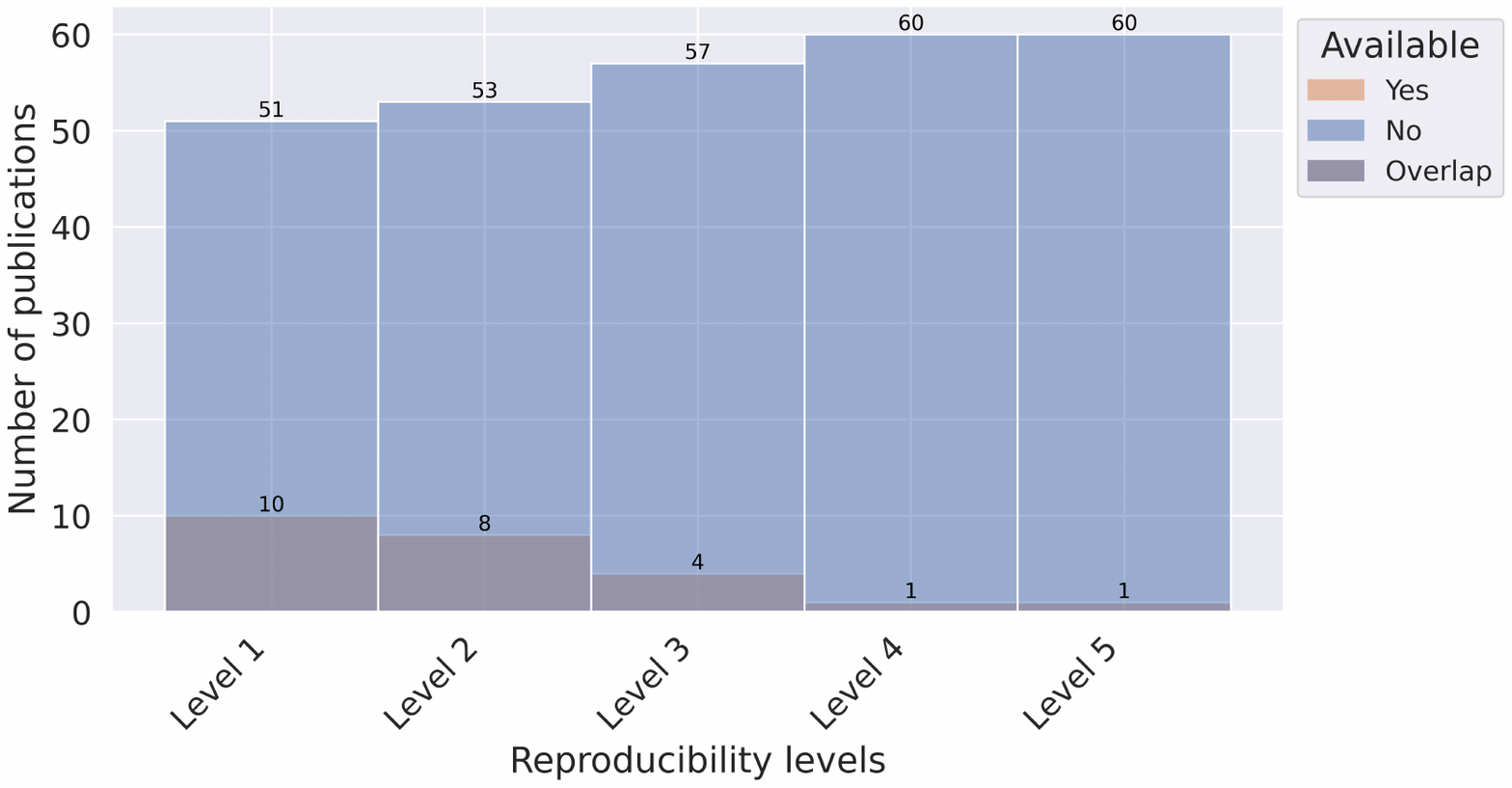}
    \caption{\centering Bar plot indicating the number of publications satisfying the different levels of reproducibility}
    \label{fig:repro-levels}
\end{figure}
\subsection*{Reproducibility levels of publications}
Per the definitions of the reproducibility levels described in Section~\ref{sec:methodology}, Levels 1 and 5 are the lowest and highest, respectively. According to Figure~\ref{fig:repro-levels}, only one publication fulfils the highest reproducibility level, while ten publications meet the lowest.

\begin{figure}
    \centering
    \begin{subfigure}{0.48\textwidth}
        \centering
        \includegraphics[width=\linewidth]{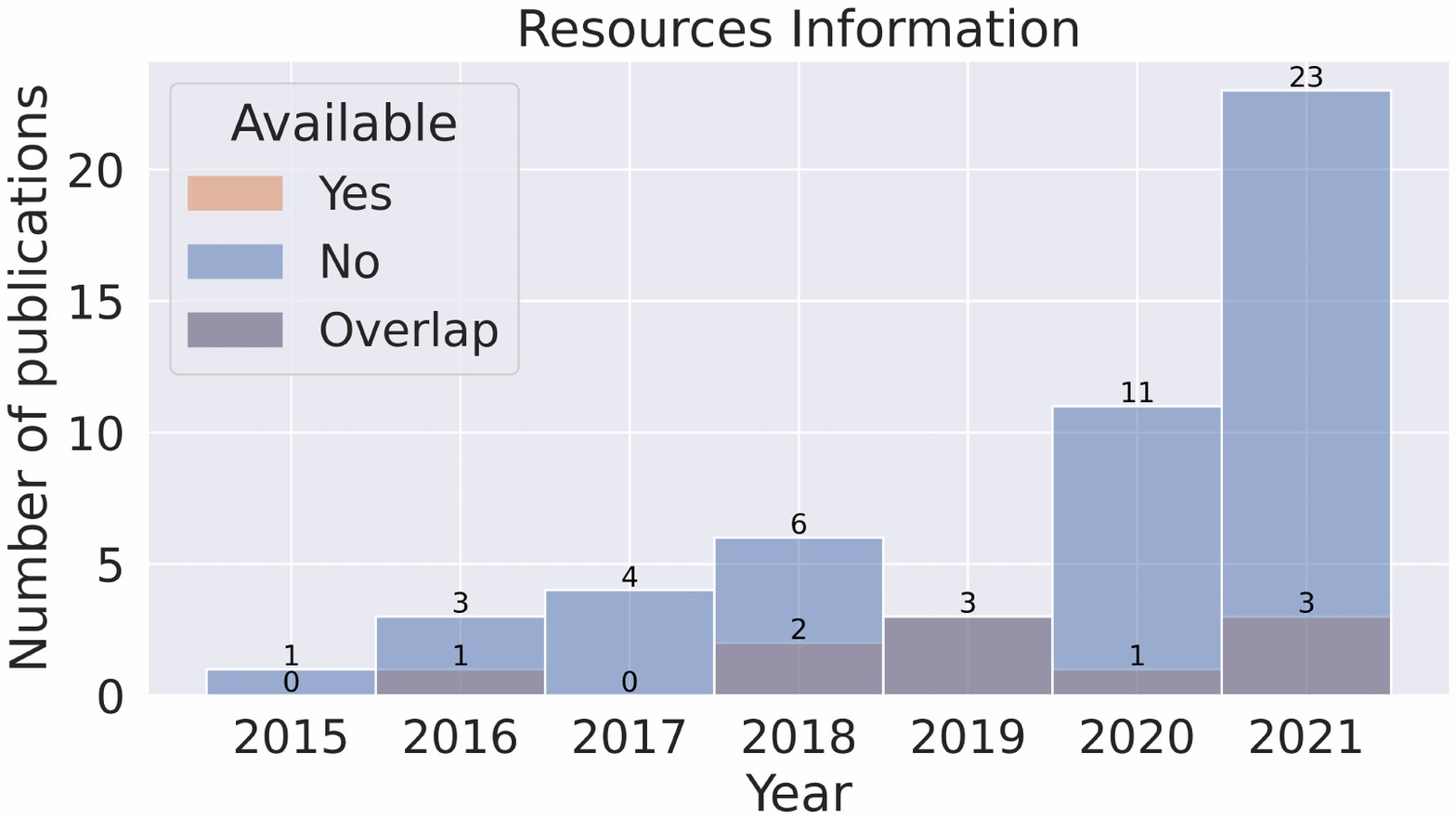}
        \caption{\centering Number of papers meeting criteria for the category 'Resources' according to year}
        \label{fig:temp-subplot1}
    \end{subfigure}
    \hfill
    \begin{subfigure}{0.48\textwidth}l 
        \centering
        \includegraphics[width=\linewidth]{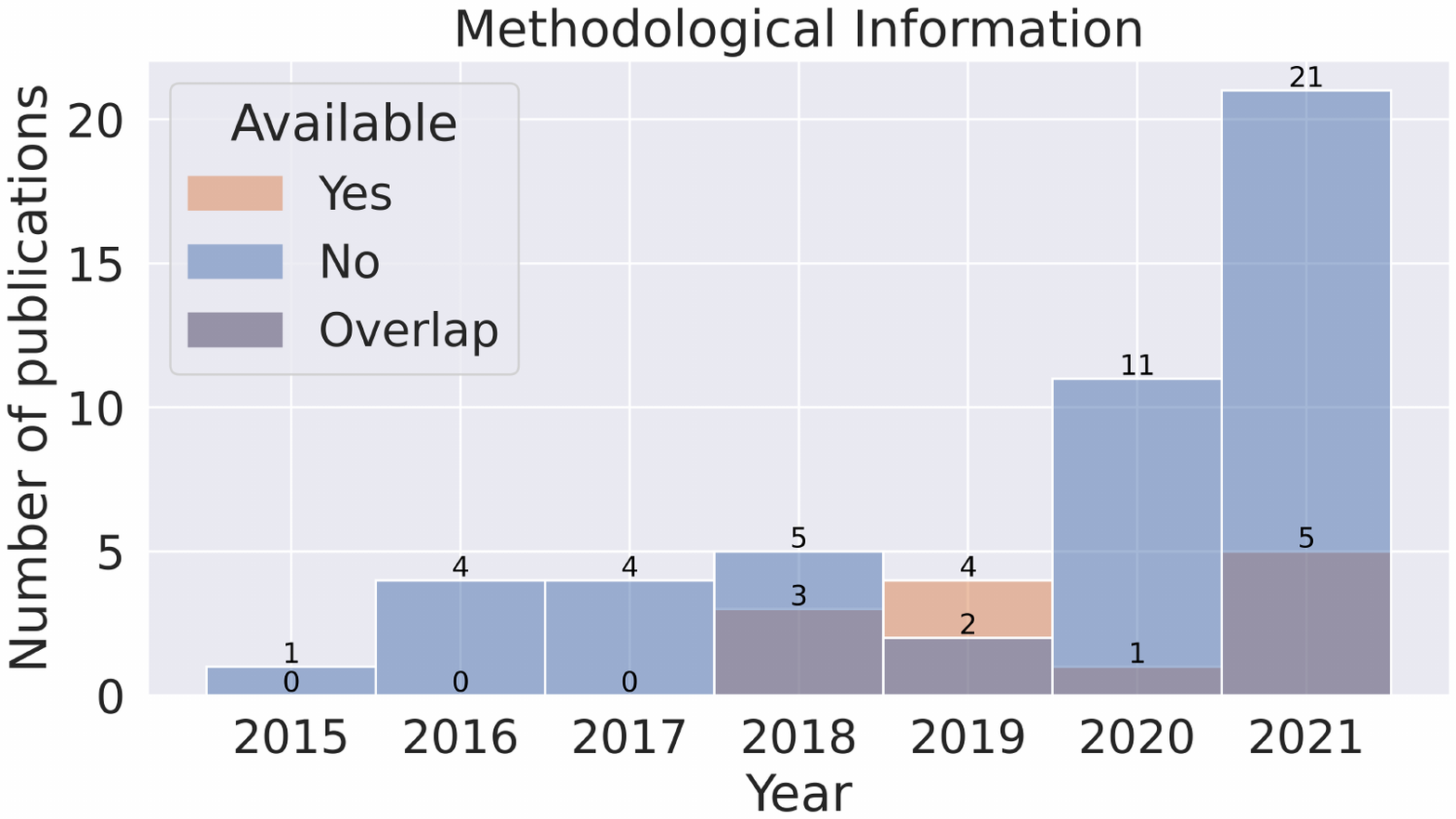}
        \caption{\centering Number of papers meeting criteria for the category 'Methodological information' according to year}
        \label{fig:temp-subplot2}
    \end{subfigure}
    \centering 
    \begin{subfigure}{0.48\textwidth}
        \centering
        \includegraphics[width=\linewidth]{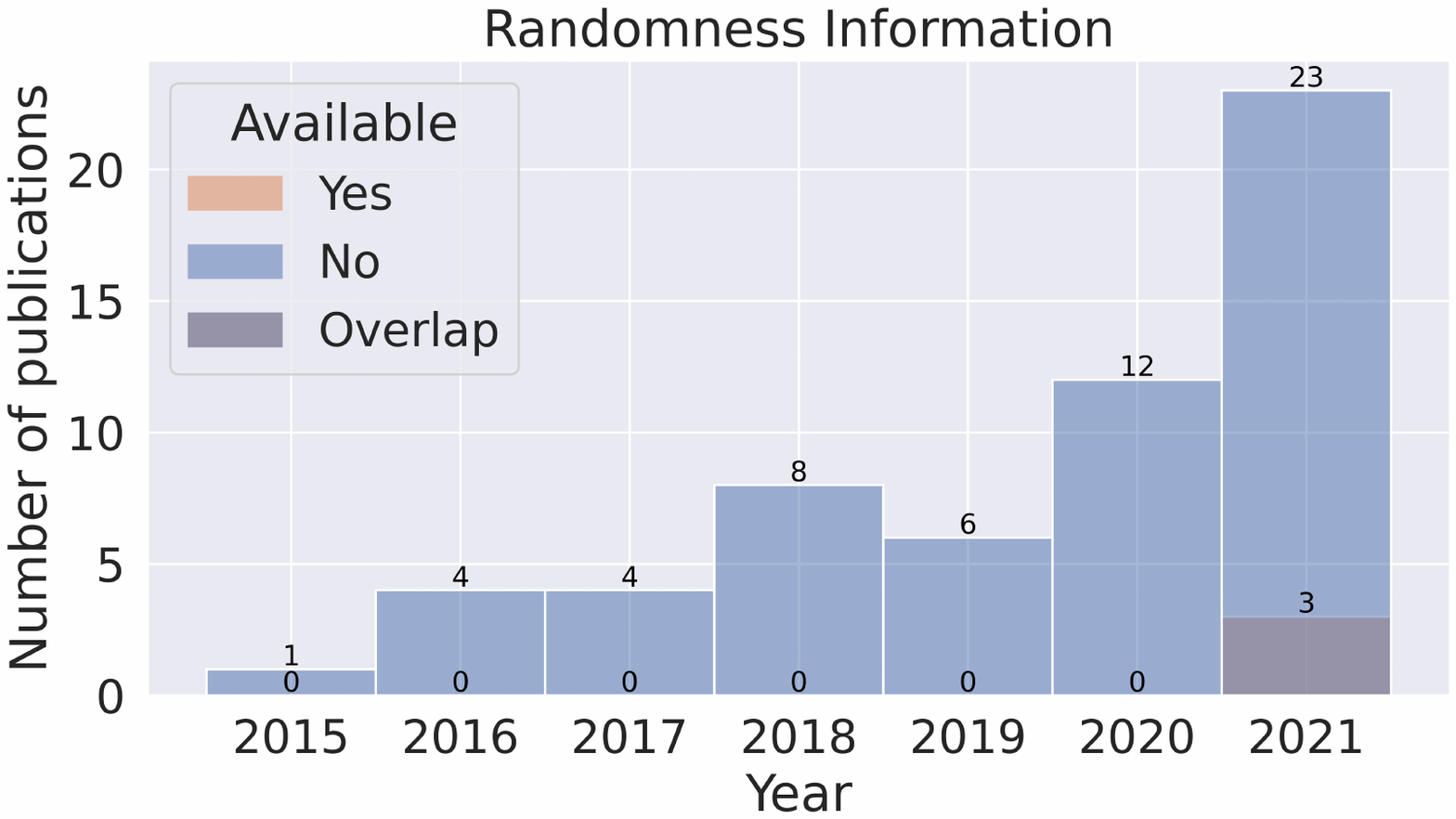}
        \caption{\centering Number of papers meeting criteria for the category 'Randomness' according to year}
        \label{fig:temp-subplot3}
    \end{subfigure}
    \hfill
    \begin{subfigure}{0.48\textwidth}
        \centering
        \includegraphics[width=\linewidth]{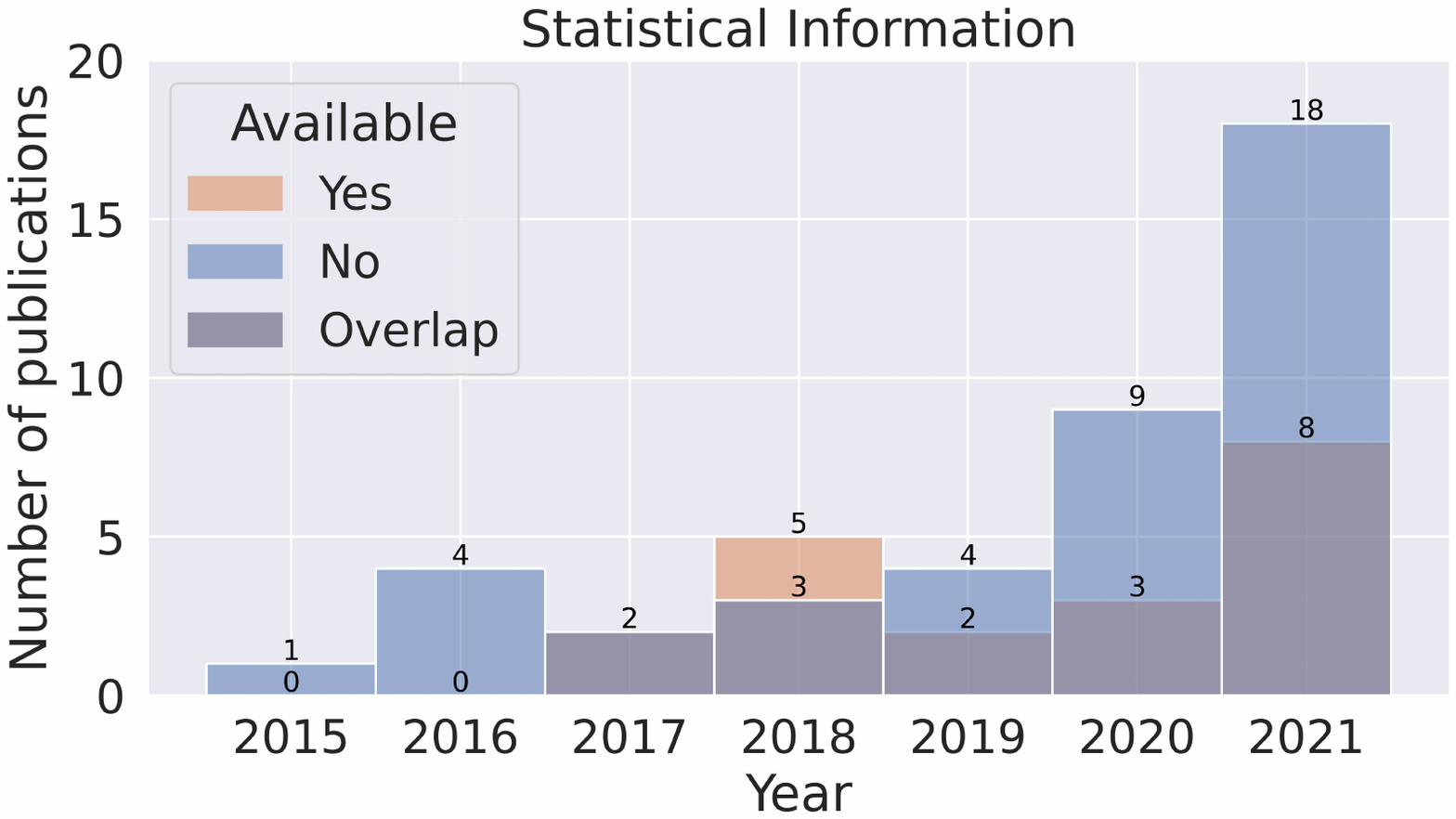}
        \caption{\centering Number of papers meeting criteria for the category 'Statistical consideration' according to year}
        \label{fig:temp-subplot4}
    \end{subfigure}
    \caption{\centering
Number of papers meeting criteria for the four categories 1) Resources 2) Methodological information 3) Randomness 4) Statistical consideration for selected research publications by year}
    \label{fig:temp-binary-responses}
\end{figure}

\subsection*{Reproducibility status of publications by year}
The number of publications that meet the defined reproducibility criteria follows a linear trend with respect to year (Figure~\ref{fig:temp-binary-responses}).

\section*{Discussion}
\label{sec:discussion}
In the Biodiversity field, deep-learning methods are becoming part of many studies that run large-scale experiments. These gave us the opportunity to orchestrate and extract the binary responses of 10 reproducibility variables from 61 publications (Table~\ref{Tab:Def} and Table~\ref{Tab:Bin-res}). As a result, we recorded 610 total responses: 353 were positive, and 257 were negative. All the positive responses were dominated by four variables (230 responses): 1) Open source frameworks or environment, 2) Model architecture, 3) Methods, and 4) Evaluation metrics and the negative responses were dominated by only one variable: Randomness.

Most of the publications that employed deep learning models use open-source frameworks or environments like Tensorflow and PyTorch with the programming language Python/R, and they also provide model architectures either as a figure/table in the publication or described in the text with respective citations. Some of the publications used licensed programming languages like Matlab, comparatively, it is negligible. We looked for high-level information on the whole deep-learning pipeline in methods, all the publications provided compact information. 

Most of the publications use more than one metric to evaluate their models, for example, when it comes to regression tasks, they use $R^2$ score along with Root Mean Square Error (RMSE) or Mean Absolute Error (MAE) or Loss of the model, etc. The information from the publications is in compliance with reproducible workflow guidelines for the positively dominated variables.

We found datasets only in 29 publications; in the other 32 publications, there was no tangential information about the dataset. There were some cases where the authors linked to some data-providing websites to find the data that was used in the publication, but those websites will change over time and finding the exact data that was used in the publication is not possible without providing persistent identifiers of the respective data points.

Source code is one of the fundamental variables of reproducibility. However, a little more than 25 \% of the publications provided their source code.

In 14 publications, authors have provided specifications about both software and hardware. Without specific information about the hardware and software, the reproducibility results will change because the random generators work differently with different hardware and software changes with each version. 

Hyperparameters are the values that are chosen to control the model learning process. This means with each set of hyperparameters, the model will provide a varied range of results. However, information about basic hyperparameters (epochs, learning rate, optimizer and loss function) was missing from 21 publications. Reproducing the results of these 21 publications is unfeasible due to the missing essential hyperparameter information.


Averaging results is also an important aspect of reproducibility, after each training process, results will change slightly because of the certain random initializations through the deep learning pipeline. However, in our study, 40 publications didn't report the multiple training results. 


We opted for manually updating the variables by going through each publication and extracting the required information as a binary response because we did not find a system or technique that could automatically extract the required variable information from a publication. Since we are extracting the information about variables from publications manually, it is only possible to work with a small dataset, which is also the limitation of this study. 

Due to recent developments with Large Language Models (LLMs), we are considering extracting the reproducible variables information from publications using LLMs from the year 2022 onwards \citep{ahmed2023reproducible, kommineni2024human}. This will allow us to implement our analysis on large-scale publications. 

\section*{Conclusion}
\label{sec:conclusion}
In this paper, we presented our pipeline for assessing the reproducibility of deep learning methods in biodiversity research.
Inspired by the current state of the art, we established a comprehensive set of ten variables, categorized into four distinct groups, to effectively quantify the reproducibility of DL empirical research.
Based on the defined categories, we documented the availability of each variable across 61 selected publications over the period from 2015 to 2021.
From the total 610 responses for the 10 variables,  57.9\% show the availability of the variables in the publication, while the remaining 42.1\% are primarily characterized by the absence of randomness-related information.
The highest and lowest reproducibility levels are satisfied by only one and ten publications, respectively.
Given the use of deep learning to advance biodiversity research, improving reproducibility of the DL methods is crucial.
Considering the limitations of the manual approach and the relatively small dataset analyzed until 2021, our future endeavors will focus on implementing a semi-automatic approach that leverages Large Language models for extracting information on reproducible variables from publications.

\section*{Data availability statement}
\label{sec:dataavailstatement}
The data and the code used to extract and analyse the reproducibility information of Deep Learning methods from publications in the Biodiversity domain is publicly available: \url{https://github.com/fusion-jena/Reproduce-DLmethods-Biodiv}

\newpage
\begin{appendices}
\fontsize{10}{12}\selectfont
\renewcommand{\arraystretch}{2}
\begin{longtable}{p{4.5cm}*{10}{c}}
\caption{Binary responses of different reproducibility variables. 'y' denotes the presence of variable information, while 'n' signifies the absence of variable information.}
\label{Tab:Bin-res} \\

\toprule
Publication & V1 & V2 & V3 & V4 & V5 & V6 & V7 & R & S1 & S2 \\
\midrule
\endfirsthead

\multicolumn{11}{c}{{\tablename\ \thetable{} -- Continued from previous page}} \\
\toprule
Publication & V1 & V2 & V3 & V4 & V5 & V6 & V7 & R & S1 & S2 \\
\midrule
\endhead

\bottomrule
\multicolumn{11}{r}{{Continued on next page}} \\
\endfoot

\bottomrule
\endlastfoot

\cite{klein2015deep} & n & n & y & n & n & y & n & n & n & n \\
\cite{Khalighifar2021-ad} & n & n & y & y & n & y & n & n & n & y \\
\cite{Choe2021-ar} & n & n & y & y & y & y & y & n & y & y \\
\cite{Mahmood2016-nx} & y & n & y & y & n & y & n & n & n & y \\
\cite{Younis2020-pz} & y & y & y & y & y & y & y & n & y & y \\
\cite{Schwartz2021-jx} & y & y & y & y & y & y & y & n & y & y \\
\cite{potamitis2016deep} & n & n & y & y & n & y & n & n & n & n \\
\cite{Chen2020-iv} & n & n & y & y & n & y & n & n & n & y \\
\cite{Fujisawa2021-gw} & n & n & y & y & n & y & n & n & y & y \\
\cite{Weinstein2018-kj} & n & y & y & y & y & y & y & n & n & y \\
\cite{Chalmers2021-sn} & n & n & y & y & y & y & n & n & n & y \\
\cite{Guirado2019-rf} & y & n & y & y & y & y & y & n & y & y \\
\cite{Zualkernan2020-hb} & n & n & y & y & n & y & y & n & n & y \\
\cite{Villon2018-vk} & n & n & y & y & n & y & y & n & y & y \\
\cite{Fairbrass2019-ep} & y & y & y & y & y & y & y & n & n & y \\
\cite{Weinstein2019-wl} & y & y & y & y & y & y & y & n & n & y \\
\cite{Hu2020-zz} & n & n & y & y & n & y & n & n & n & y \\
\cite{Alshahrani2021-cv} & y & n & y & y & n & y & y & n & y & y \\
\cite{Villon2020-rn} & n & n & y & y & n & y & n & n & y & y \\
\cite{Botella2018-xa} & y & n & y & y & n & y & y & n & y & y \\
\cite{Salamon2017-ap} & y & n & y & y & n & y & y & n & y & y \\
\cite{Mac_Aodha2018-kw} & y & y & y & y & y & y & y & n & y & y \\
\cite{Schindler2021-nw} & n & n & y & y & n & y & y & n & n & y \\
\cite{Rakshit2018-ae} & n & n & y & y & n & y & y & n & n & y \\
\cite{Bjerge2021-la} & n & y & y & y & n & y & y & y & y & y \\
\cite{Guirado2017-ao} & n & n & y & y & n & y & n & n & n & y \\
\cite{Hussein2021-vg} & n & n & y & y & n & y & y & n & n & y \\
\cite{Rammer2019-lx} & y & y & y & y & y & y & y & n & n & y \\
\cite{Anand2021-uh} & n & n & n & y & n & y & n & n & n & y \\
\cite{Zizka2021-qc} & y & y & y & y & n & y & y & y & n & n \\
\cite{Miele2020-bb} & n & n & y & y & n & y & n & n & y & n \\
\cite{Demertzis2018-sw} & n & n & y & y & n & y & n & n & y & y \\
\cite{Malerba2021-ij} & y & n & y & y & n & y & y & n & n & n \\
\cite{Huang2021-av} & n & n & y & y & n & y & y & n & y & y \\
\cite{Campos-Taberner2020-yt} & y & n & n & y & n & y & n & n & n & y \\
\cite{Rousset2021-bj} & y & n & y & y & n & y & y & n & n & y \\
\cite{Lopez-Jimenez2019-rh} & y & n & y & y & n & y & y & n & n & y \\
\cite{Schuettpelz2017-xc} & n & y & y & y & n & y & y & n & n & y \\
\cite{Schiller2021-nt} & y & y & y & y & y & y & y & y & y & y \\
\cite{Heredia2017-ys} & n & y & y & y & n & y & y & n & y & y \\
\cite{Martins2021-ab} & y & n & y & y & y & y & y & n & n & y \\
\cite{Browning2018-go} & y & n & y & y & n & y & n & n & y & y \\
\cite{Guirado2020-qi} & y & n & y & y & n & y & y & n & n & y \\
\cite{Ayhan2020-vc} & n & n & y & y & n & y & y & n & n & y \\
\cite{Loddo2021-ym} & n & n & y & y & n & y & n & n & y & y \\
\cite{Jamil2021-we} & y & n & y & y & n & y & y & n & n & y \\
\cite{Neves2021-sx} & n & n & y & y & n & y & n & n & n & y \\
\cite{Jin2021-ii} & y & n & y & n & n & y & n & n & n & y \\
\cite{Mohanty2016-vk} & y & y & y & y & n & y & y & n & n & y \\
\cite{Xie2019-gn} & y & n & y & y & n & y & y & n & y & y \\
\cite{Tian2020-mw} & n & n & n & n & n & y & n & n & y & y \\
\cite{Gimenez2021-qt} & n & y & y & y & n & y & y & n & n & y \\
\cite{Ortega_Adarme2020-qj} & y & n & y & y & n & y & y & n & n & y \\
\cite{boer2018taxonomic} & y & y & y & y & y & y & y & n & n & n \\
\cite{Dunker2021-ul} & n & n & y & y & n & y & n & n & n & y \\
\cite{Villon2016-yu} & n & n & y & y & n & y & n & n & n & y \\
\cite{Dyrmann2021-or} & n & n & y & y & n & y & y & n & n & y \\
\cite{Arruda2021-cx} & y & n & y & n & n & y & y & n & n & n \\
\cite{Kislov2020-nc} & y & n & y & y & n & y & y & n & n & y \\
\cite{Hussein2021-et} & y & n & y & y & n & y & y & n & n & y \\
\cite{Becker2021-ac} & n & y & y & y & y & y & y & n & n & y \\
\bottomrule
\end{longtable}

\end{appendices}








\bibliography{main}

\begin{thebibliography}{}

\bibitem[Abdelmageed et~al., 2022]{abdelmageed2022biodivnere}
Abdelmageed, N., L{\"o}ffler, F., Feddoul, L., Algergawy, A., Samuel, S.,
  Gaikwad, J., Kazem, A., and K{\"o}nig-Ries, B. (2022).
\newblock Biodivnere: Gold standard corpora for named entity recognition and
  relation extraction in the biodiversity domain.
\newblock {\em Biodiversity Data Journal}, 10.

\bibitem[Ahmed et~al., 2023]{ahmed2023reproducible}
Ahmed, W., Kommineni, V.~K., Koenig-ries, B., and Samuel, S. (2023).
\newblock How reproducible are the results gained with the help of deep
  learning methods in biodiversity research?
\newblock {\em Biodiversity Information Science and Standards}, 7.

\bibitem[Alshahrani et~al., 2021]{Alshahrani2021-cv}
Alshahrani, H.~M., Al-Wesabi, F.~N., Al~Duhayyim, M., Nemri, N., Kadry, S., and
  Alqaralleh, B. A.~Y. (2021).
\newblock An automated deep learning based satellite imagery analysis for
  ecology management.
\newblock {\em Ecol. Inform.}, 66(101452):101452.

\bibitem[Anand et~al., 2021]{Anand2021-uh}
Anand, A., Pandey, M.~K., Srivastava, P.~K., Gupta, A., and Khan, M.~L. (2021).
\newblock Integrating multi-sensors data for species distribution mapping using
  deep learning and envelope models.
\newblock {\em Remote Sens. (Basel)}, 13(16):3284.

\bibitem[Arruda et~al., 2021]{Arruda2021-cx}
Arruda, V. L.~S., Piontekowski, V.~J., Alencar, A., Pereira, R.~S., and
  Matricardi, E. A.~T. (2021).
\newblock An alternative approach for mapping burn scars using landsat imagery,
  google earth engine, and deep learning in the brazilian savanna.
\newblock {\em Remote Sens. Appl. Soc. Environ.}, 22(100472):100472.

\bibitem[August et~al., 2020]{AUGUST2020100116}
August, T.~A., Pescott, O.~L., Joly, A., and Bonnet, P. (2020).
\newblock {AI} naturalists might hold the key to unlocking biodiversity data in
  social media imagery.
\newblock {\em Patterns}, 1(7):100116.

\bibitem[Ayhan et~al., 2020]{Ayhan2020-vc}
Ayhan, B., Kwan, C., Budavari, B., Kwan, L., Lu, Y., Perez, D., Li, J.,
  Skarlatos, D., and Vlachos, M. (2020).
\newblock Vegetation detection using deep learning and conventional methods.
\newblock {\em Remote Sens. (Basel)}, 12(15):2502.

\bibitem[Becker et~al., 2021]{Becker2021-ac}
Becker, A., Russo, S., Puliti, S., Lang, N., Schindler, K., and Wegner, J.~D.
  (2021).
\newblock Country-wide retrieval of forest structure from optical and {SAR}
  satellite imagery with deep ensembles.

\bibitem[Bjerge et~al., 2021]{Bjerge2021-la}
Bjerge, K., Nielsen, J.~B., Sepstrup, M.~V., Helsing-Nielsen, F., and H{\o}ye,
  T.~T. (2021).
\newblock An automated light trap to monitor moths (lepidoptera) using computer
  vision-based tracking and deep learning.
\newblock {\em Sensors (Basel)}, 21(2):343.

\bibitem[Boer and Vos, 2018]{boer2018taxonomic}
Boer, M.~J. and Vos, R.~A. (2018).
\newblock Taxonomic classification of ants (formicidae) from images using deep
  learning.
\newblock {\em bioRxiv}, page 407452.

\bibitem[Botella et~al., 2018]{Botella2018-xa}
Botella, C., Joly, A., Bonnet, P., Monestiez, P., and Munoz, F. (2018).
\newblock A deep learning approach to species distribution modelling.
\newblock In {\em Multimedia Tools and Applications for Environmental \&
  Biodiversity Informatics}, pages 169--199. Springer International Publishing,
  Cham.

\bibitem[Browning et~al., 2018]{Browning2018-go}
Browning, E., Bolton, M., Owen, E., Shoji, A., Guilford, T., and Freeman, R.
  (2018).
\newblock Predicting animal behaviour using deep learning: {GPS} data alone
  accurately predict diving in seabirds.
\newblock {\em Methods Ecol. Evol.}, 9(3):681--692.

\bibitem[Campos-Taberner et~al., 2020]{Campos-Taberner2020-yt}
Campos-Taberner, M., Garc{\'\i}a-Haro, F.~J., Mart{\'\i}nez, B.,
  Izquierdo-Verdiguier, E., Atzberger, C., Camps-Valls, G., and Gilabert, M.~A.
  (2020).
\newblock Understanding deep learning in land use classification based on
  sentinel-2 time series.
\newblock {\em Sci. Rep.}, 10(1):17188.

\bibitem[Chalmers et~al., 2021]{Chalmers2021-sn}
Chalmers, C., Fergus, P., Wich, S., and Longmore, S.~N. (2021).
\newblock Modelling animal biodiversity using acoustic monitoring and deep
  learning.
\newblock In {\em 2021 International Joint Conference on Neural Networks
  ({IJCNN})}. IEEE.

\bibitem[Chen et~al., 2020]{Chen2020-iv}
Chen, X., Zhao, J., Chen, Y.-H., Zhou, W., and Hughes, A.~C. (2020).
\newblock Automatic standardized processing and identification of tropical bat
  calls using deep learning approaches.
\newblock {\em Biol. Conserv.}, 241(108269):108269.

\bibitem[Choe et~al., 2021]{Choe2021-ar}
Choe, H., Chi, J., and Thorne, J.~H. (2021).
\newblock Mapping potential plant species richness over large areas with deep
  learning, {MODIS}, and species distribution models.
\newblock {\em Remote Sens. (Basel)}, 13(13):2490.

\bibitem[Christin et~al., 2019]{christin2019applications}
Christin, S., Hervet, {\'E}., and Lecomte, N. (2019).
\newblock Applications for deep learning in ecology.
\newblock {\em Methods in Ecology and Evolution}, 10(10):1632--1644.

\bibitem[Cohen, 1960]{cohen1960coefficient}
Cohen, J. (1960).
\newblock A coefficient of agreement for nominal scales.
\newblock {\em Educational and psychological measurement}, 20(1):37--46.

\bibitem[Demertzis et~al., 2018]{Demertzis2018-sw}
Demertzis, K., Iliadis, L.~S., and Anezakis, V.-D. (2018).
\newblock Extreme deep learning in biosecurity: the case of machine hearing for
  marine species identification.
\newblock {\em J. Inf. Telecommun.}, 2(4):492--510.

\bibitem[Dunker et~al., 2021]{Dunker2021-ul}
Dunker, S., Motivans, E., Rakosy, D., Boho, D., M{\"a}der, P., Hornick, T., and
  Knight, T.~M. (2021).
\newblock Pollen analysis using multispectral imaging flow cytometry and deep
  learning.
\newblock {\em New Phytol.}, 229(1):593--606.

\bibitem[Dyrmann et~al., 2021]{Dyrmann2021-or}
Dyrmann, M., Mortensen, A.~K., Linneberg, L., H{\o}ye, T.~T., and Bjerge, K.
  (2021).
\newblock Camera assisted roadside monitoring for invasive alien plant species
  using deep learning.
\newblock {\em Sensors (Basel)}, 21(18):6126.

\bibitem[El-Amir and Hamdy, 2020]{el2020deep}
El-Amir, H. and Hamdy, M. (2020).
\newblock Deep learning pipeline.
\newblock {\em Apress: Berkeley, CA, USA}.

\bibitem[Fairbrass et~al., 2019]{Fairbrass2019-ep}
Fairbrass, A.~J., Firman, M., Williams, C., Brostow, G.~J., Titheridge, H., and
  Jones, K.~E. (2019).
\newblock {CityNet---Deep} learning tools for urban ecoacoustic assessment.
\newblock {\em Methods Ecol. Evol.}, 10(2):186--197.

\bibitem[Feng et~al., 2019]{feng2019checklist}
Feng, X., Park, D.~S., Walker, C., Peterson, A.~T., Merow, C., and Pape{\c{s}},
  M. (2019).
\newblock A checklist for maximizing reproducibility of ecological niche
  models.
\newblock {\em Nature Ecology \& Evolution}, 3(10):1382--1395.

\bibitem[Fujisawa et~al., 2021]{Fujisawa2021-gw}
Fujisawa, T., Noguerales, V., Meramveliotakis, E., Papadopoulou, A., and
  Vogler, A.~P. (2021).
\newblock Image-based taxonomic classification of bulk biodiversity samples
  using deep learning and domain adaptation.

\bibitem[Gimenez et~al., 2021]{Gimenez2021-qt}
Gimenez, O., Kervellec, M., Fanjul, J.-B., Chaine, A., Marescot, L., Bollet,
  Y., and Duchamp, C. (2021).
\newblock Trade-off between deep learning for species identification and
  inference about predator-prey co-occurrence: Reproducible {R} workflow
  integrating models in computer vision and ecological statistics.

\bibitem[Goodman et~al., 2016]{goodman2016what}
Goodman, S.~N., Fanelli, D., and Ioannidis, J. P.~A. (2016).
\newblock What does research reproducibility mean?
\newblock {\em Science Translational Medicine}, 8(341):341ps12--341ps12.

\bibitem[GPAI, 2022]{gpai2022biodiversity}
GPAI (2022).
\newblock Biodiversity and artificial intelligence, opportunities and
  recommendations report.

\bibitem[Guirado et~al., 2020]{Guirado2020-qi}
Guirado, E., Alcaraz-Segura, D., Cabello, J., Puertas-Ru{\'\i}z, S., Herrera,
  F., and Tabik, S. (2020).
\newblock Tree cover estimation in global drylands from space using deep
  learning.
\newblock {\em Remote Sens. (Basel)}, 12(3):343.

\bibitem[Guirado et~al., 2017]{Guirado2017-ao}
Guirado, E., Tabik, S., Alcaraz-Segura, D., Cabello, J., and Herrera, F.
  (2017).
\newblock {Deep-Learning} convolutional neural networks for scattered shrub
  detection with google earth imagery.

\bibitem[Guirado et~al., 2019]{Guirado2019-rf}
Guirado, E., Tabik, S., Rivas, M.~L., Alcaraz-Segura, D., and Herrera, F.
  (2019).
\newblock Whale counting in satellite and aerial images with deep learning.
\newblock {\em Sci. Rep.}, 9(1):14259.

\bibitem[Gundersen et~al., 2022]{gundersen2022machine}
Gundersen, O.~E., Shamsaliei, S., and Isdahl, R.~J. (2022).
\newblock Do machine learning platforms provide out-of-the-box reproducibility?
\newblock {\em Future Generation Computer Systems}, 126:34--47.

\bibitem[Heil et~al., 2021]{heil2021reproducibility}
Heil, B.~J., Hoffman, M.~M., Markowetz, F., Lee, S.-I., Greene, C.~S., and
  Hicks, S.~C. (2021).
\newblock Reproducibility standards for machine learning in the life sciences.
\newblock {\em Nature Methods}, 18(10):1132--1135.

\bibitem[Heredia, 2017]{Heredia2017-ys}
Heredia, I. (2017).
\newblock Large-scale plant classification with deep neural networks.
\newblock In {\em Proceedings of the Computing Frontiers Conference}, New York,
  NY, USA. ACM.

\bibitem[Hu et~al., 2020]{Hu2020-zz}
Hu, J., Huang, W., Su, Y., Liu, Y., and Xiao, P. (2020).
\newblock {BatNet++}: A robust deep learning-based predicting models for calls
  recognition.
\newblock In {\em 2020 5th International Conference on Smart Grid and
  Electrical Automation ({ICSGEA})}. IEEE.

\bibitem[Huang and Basanta, 2021]{Huang2021-av}
Huang, Y.-P. and Basanta, H. (2021).
\newblock Recognition of endemic bird species using deep learning models.
\newblock {\em IEEE Access}, 9:102975--102984.

\bibitem[Hussein et~al., 2021a]{Hussein2021-vg}
Hussein, B.~R., Malik, O.~A., Ong, W.-H., and Slik, J. W.~F. (2021a).
\newblock Automated extraction of phenotypic leaf traits of individual intact
  herbarium leaves from herbarium specimen images using deep learning based
  semantic segmentation.
\newblock {\em Sensors (Basel)}, 21(13):4549.

\bibitem[Hussein et~al., 2021b]{Hussein2021-et}
Hussein, B.~R., Malik, O.~A., Ong, W.-H., and Slik, J. W.~F. (2021b).
\newblock Reconstruction of damaged herbarium leaves using deep learning
  techniques for improving classification accuracy.
\newblock {\em Ecol. Inform.}, 61(101243):101243.

\bibitem[Jamil et~al., 2021]{Jamil2021-we}
Jamil, S., Rahman, M., and Haider, A. (2021).
\newblock Bag of features ({BoF}) based deep learning framework for bleached
  corals detection.
\newblock {\em Big Data Cogn. Comput.}, 5(4):53.

\bibitem[Jin et~al., 2021]{Jin2021-ii}
Jin, L., Yu, J., Yuan, X., and Du, X. (2021).
\newblock Fish classification using {DNA} barcode sequences through deep
  learning method.
\newblock {\em Symmetry (Basel)}, 13(9):1599.

\bibitem[Khalighifar et~al., 2021]{Khalighifar2021-ad}
Khalighifar, A., Brown, R.~M., Goyes~Vallejos, J., and Peterson, A.~T. (2021).
\newblock Deep learning improves acoustic biodiversity monitoring and new
  candidate forest frog species identification (genus platymantis) in the
  {P}hilippines.
\newblock {\em Biodivers. Conserv.}, 30(3):643--657.

\bibitem[Kislov and Korznikov, 2020]{Kislov2020-nc}
Kislov, D.~E. and Korznikov, K.~A. (2020).
\newblock Automatic windthrow detection using very-high-resolution satellite
  imagery and deep learning.
\newblock {\em Remote Sens. (Basel)}, 12(7):1145.

\bibitem[Klein et~al., 2015]{klein2015deep}
Klein, D.~J., McKown, M.~W., and Tershy, B.~R. (2015).
\newblock Deep learning for large scale biodiversity monitoring.
\newblock In {\em Bloomberg Data for Good Exchange Conference}.

\bibitem[Kommineni et~al., 2024]{kommineni2024human}
Kommineni, V.~K., K{\"o}nig-Ries, B., and Samuel, S. (2024).
\newblock From human experts to machines: An llm supported approach to ontology
  and knowledge graph construction.
\newblock {\em arXiv preprint arXiv:2403.08345}.

\bibitem[Leonelli, 2018]{leonelli2018rethinking}
Leonelli, S. (2018).
\newblock Rethinking reproducibility as a criterion for research quality.
\newblock In {\em Including a symposium on Mary Morgan: curiosity, imagination,
  and surprise}, volume~36, pages 129--146. Emerald Publishing Limited.

\bibitem[Loddo et~al., 2021]{Loddo2021-ym}
Loddo, A., Loddo, M., and Di~Ruberto, C. (2021).
\newblock A novel deep learning based approach for seed image classification
  and retrieval.
\newblock {\em Comput. Electron. Agric.}, 187(106269):106269.

\bibitem[L{\'o}pez-Jim{\'e}nez et~al., 2019]{Lopez-Jimenez2019-rh}
L{\'o}pez-Jim{\'e}nez, E., Vasquez-Gomez, J.~I., Sanchez-Acevedo, M.~A.,
  Herrera-Lozada, J.~C., and Uriarte-Arcia, A.~V. (2019).
\newblock Columnar cactus recognition in aerial images using a deep learning
  approach.
\newblock {\em Ecol. Inform.}, 52:131--138.

\bibitem[Mac~Aodha et~al., 2018]{Mac_Aodha2018-kw}
Mac~Aodha, O., Gibb, R., Barlow, K.~E., Browning, E., Firman, M., Freeman, R.,
  Harder, B., Kinsey, L., Mead, G.~R., Newson, S.~E., Pandourski, I., Parsons,
  S., Russ, J., Szodoray-Paradi, A., Szodoray-Paradi, F., Tilova, E., Girolami,
  M., Brostow, G., and Jones, K.~E. (2018).
\newblock Bat detective---deep learning tools for bat acoustic signal
  detection.
\newblock {\em PLoS Comput. Biol.}, 14(3):e1005995.

\bibitem[Mahmood et~al., 2016]{Mahmood2016-nx}
Mahmood, A., Bennamoun, M., An, S., Sohel, F., Boussaid, F., Hovey, R.,
  Kendrick, G., and Fisher, R.~B. (2016).
\newblock Automatic annotation of coral reefs using deep learning.
\newblock In {\em {OCEANS} 2016 {MTS/IEEE} Monterey}. IEEE.

\bibitem[Malerba et~al., 2021]{Malerba2021-ij}
Malerba, M.~E., Wright, N., and Macreadie, P.~I. (2021).
\newblock A continental-scale assessment of density, size, distribution and
  historical trends of farm dams using deep learning convolutional neural
  networks.
\newblock {\em Remote Sens. (Basel)}, 13(2):319.

\bibitem[Martins et~al., 2021]{Martins2021-ab}
Martins, J. A.~C., Nogueira, K., Osco, L.~P., Gomes, F. D.~G., Furuya, D.
  E.~G., Gon{\c c}alves, W.~N., Sant'Ana, D.~A., Ramos, A. P.~M., Liesenberg,
  V., dos Santos, J.~A., de~Oliveira, P. T.~S., and Junior, J.~M. (2021).
\newblock Semantic segmentation of tree-canopy in urban environment with
  pixel-wise deep learning.
\newblock {\em Remote Sens. (Basel)}, 13(16):3054.

\bibitem[Miele et~al., 2020]{Miele2020-bb}
Miele, V., Dussert, G., Cucchi, T., and Renaud, S. (2020).
\newblock Deep learning for species identification of modern and fossil rodent
  molars.

\bibitem[Mohanty et~al., 2016]{Mohanty2016-vk}
Mohanty, S.~P., Hughes, D.~P., and Salath{\'e}, M. (2016).
\newblock Using deep learning for image-based plant disease detection.
\newblock {\em Front. Plant Sci.}, 7:1419.

\bibitem[Neves et~al., 2021]{Neves2021-sx}
Neves, A.~K., K{\"o}rting, T.~S., Fonseca, L. M.~G., Soares, A.~R.,
  Girolamo-Neto, C.~D., and Heipke, C. (2021).
\newblock Hierarchical mapping of brazilian savanna (cerrado) physiognomies
  based on deep learning.
\newblock {\em J. Appl. Remote Sens.}, 15(04).

\bibitem[Norouzzadeh et~al., 2018]{norouzzadeh2018automatically}
Norouzzadeh, M.~S., Nguyen, A., Kosmala, M., Swanson, A., Palmer, M.~S.,
  Packer, C., and Clune, J. (2018).
\newblock Automatically identifying, counting, and describing wild animals in
  camera-trap images with deep learning.
\newblock {\em Proceedings of the National Academy of Sciences},
  115(25):E5716--E5725.

\bibitem[Ortega~Adarme et~al., 2020]{Ortega_Adarme2020-qj}
Ortega~Adarme, M., Queiroz~Feitosa, R., Nigri~Happ, P., Aparecido De~Almeida,
  C., and Rodrigues~Gomes, A. (2020).
\newblock Evaluation of deep learning techniques for deforestation detection in
  the brazilian amazon and cerrado biomes from remote sensing imagery.
\newblock {\em Remote Sens. (Basel)}, 12(6):910.

\bibitem[Pineau et~al., 2021]{pineau2021improving}
Pineau, J., Vincent-Lamarre, P., Sinha, K., Larivi{\`e}re, V., Beygelzimer, A.,
  d'Alch{\'e} Buc, F., Fox, E., and Larochelle, H. (2021).
\newblock Improving reproducibility in machine learning research (a report from
  the neurips 2019 reproducibility program).
\newblock {\em The Journal of Machine Learning Research}, 22(1):7459--7478.

\bibitem[Potamitis, 2016]{potamitis2016deep}
Potamitis, I. (2016).
\newblock Deep learning for detection of bird vocalisations.
\newblock {\em arXiv preprint arXiv:1609.08408}.

\bibitem[Raff, 2019]{raff2019step}
Raff, E. (2019).
\newblock A step toward quantifying independently reproducible machine learning
  research.
\newblock {\em Advances in Neural Information Processing Systems}, 32.

\bibitem[Rakshit et~al., 2018]{Rakshit2018-ae}
Rakshit, S., Debnath, S., and Mondal, D. (2018).
\newblock Identifying land patterns from satellite imagery in amazon rainforest
  using deep learning.

\bibitem[Rammer and Seidl, 2019]{Rammer2019-lx}
Rammer, W. and Seidl, R. (2019).
\newblock Harnessing deep learning in ecology: An example predicting bark
  beetle outbreaks.
\newblock {\em Front. Plant Sci.}, 10:1327.

\bibitem[Rousset et~al., 2021]{Rousset2021-bj}
Rousset, G., Despinoy, M., Schindler, K., and Mangeas, M. (2021).
\newblock Assessment of deep learning techniques for land use land cover
  classification in southern new caledonia.
\newblock {\em Remote Sens. (Basel)}, 13(12):2257.

\bibitem[Rovero et~al., 2013]{rovero2013camera}
Rovero, F., Zimmermann, F., Berzi, D., and Meek, P. (2013).
\newblock " which camera trap type and how many do i need?" a review of camera
  features and study designs for a range of wildlife research applications.
\newblock {\em Hystrix}.

\bibitem[Salamon et~al., 2017]{Salamon2017-ap}
Salamon, J., Bello, J.~P., Farnsworth, A., and Kelling, S. (2017).
\newblock Fusing shallow and deep learning for bioacoustic bird species
  classification.
\newblock In {\em 2017 {IEEE} International Conference on Acoustics, Speech and
  Signal Processing ({ICASSP})}. IEEE.

\bibitem[Samuel and König-Ries, 2021]{samuel2021understanding}
Samuel, S. and König-Ries, B. (2021).
\newblock Understanding experiments and research practices for reproducibility:
  an exploratory study.
\newblock {\em PeerJ}, 9:e11140.

\bibitem[Samuel et~al., 2021]{samuel2020machine}
Samuel, S., L{\"{o}}ffler, F., and K{\"{o}}nig{-}Ries, B. (2021).
\newblock Machine learning pipelines: Provenance, reproducibility and {FAIR}
  data principles.
\newblock In Glavic, B., Braganholo, V., and Koop, D., editors, {\em Provenance
  and Annotation of Data and Processes - 8th and 9th International Provenance
  and Annotation Workshop, {IPAW} 2020 + {IPAW} 2021, Virtual Event, July
  19-22, 2021, Proceedings}, volume 12839 of {\em Lecture Notes in Computer
  Science}, pages 226--230. Springer.

\bibitem[Schiller et~al., 2021]{Schiller2021-nt}
Schiller, C., Schmidtlein, S., Boonman, C., Moreno-Mart{\'\i}nez, A., and
  Kattenborn, T. (2021).
\newblock Deep learning and citizen science enable automated plant trait
  predictions from photographs.
\newblock {\em Sci. Rep.}, 11(1):16395.

\bibitem[Schindler and Steinhage, 2021]{Schindler2021-nw}
Schindler, F. and Steinhage, V. (2021).
\newblock Identification of animals and recognition of their actions in
  wildlife videos using deep learning techniques.
\newblock {\em Ecol. Inform.}, 61(101215):101215.

\bibitem[Schnitzer and Carson, 2016]{schnitzer2016would}
Schnitzer, S.~A. and Carson, W.~P. (2016).
\newblock Would ecology fail the repeatability test?
\newblock {\em BioScience}, 66(2):98--99.

\bibitem[Schuettpelz et~al., 2017]{Schuettpelz2017-xc}
Schuettpelz, E., Frandsen, P., Dikow, R., Brown, A., Orli, S., Peters, M.,
  Metallo, A., Funk, V., and Dorr, L. (2017).
\newblock Applications of deep convolutional neural networks to digitized
  natural history collections.
\newblock {\em Biodivers. Data J.}, 5:e21139.

\bibitem[Schwartz and Alfaro, 2021]{Schwartz2021-jx}
Schwartz, S.~T. and Alfaro, M.~E. (2021).
\newblock \textit{Sashimi}: A toolkit for facilitating high‐throughput
  organismal image segmentation using deep learning.
\newblock {\em Methods Ecol. Evol.}, 12(12):2341--2354.

\bibitem[Stark, 2018]{stark2018before}
Stark, P.~B. (2018).
\newblock Before reproducibility must come preproducibility.
\newblock {\em Nature}, 557(7706):613--614.

\bibitem[Tabak et~al., 2019]{tabak2019machine}
Tabak, M.~A., Norouzzadeh, M.~S., Wolfson, D.~W., Sweeney, S.~J., VerCauteren,
  K.~C., Snow, N.~P., Halseth, J.~M., Di~Salvo, P.~A., Lewis, J.~S., White,
  M.~D., et~al. (2019).
\newblock Machine learning to classify animal species in camera trap images:
  Applications in ecology.
\newblock {\em Methods in Ecology and Evolution}, 10(4):585--590.

\bibitem[Tatman et~al., 2018]{tatman2018practical}
Tatman, R., VanderPlas, J., and Dane, S. (2018).
\newblock A practical taxonomy of reproducibility for machine learning
  research.

\bibitem[Tian et~al., 2020]{Tian2020-mw}
Tian, J., Wang, L., Yin, D., Li, X., Diao, C., Gong, H., Shi, C., Menenti, M.,
  Ge, Y., Nie, S., Ou, Y., Song, X., and Liu, X. (2020).
\newblock Development of spectral-phenological features for deep learning to
  understand spartina alterniflora invasion.
\newblock {\em Remote Sens. Environ.}, 242(111745):111745.

\bibitem[Villon et~al., 2016]{Villon2016-yu}
Villon, S., Chaumont, M., Subsol, G., Vill{\'e}ger, S., Claverie, T., and
  Mouillot, D. (2016).
\newblock Coral reef fish detection and recognition in underwater videos by
  supervised machine learning: Comparison between deep learning and {HOG+SVM}
  methods.
\newblock In {\em Advanced Concepts for Intelligent Vision Systems}, Lecture
  notes in computer science, pages 160--171. Springer International Publishing,
  Cham.

\bibitem[Villon et~al., 2018]{Villon2018-vk}
Villon, S., Mouillot, D., Chaumont, M., Darling, E.~S., Subsol, G., Claverie,
  T., and Vill{\'e}ger, S. (2018).
\newblock A deep learning method for accurate and fast identification of coral
  reef fishes in underwater images.
\newblock {\em Ecol. Inform.}, 48:238--244.

\bibitem[Villon et~al., 2020]{Villon2020-rn}
Villon, S., Mouillot, D., Chaumont, M., Subsol, G., Claverie, T., and
  Vill{\'e}ger, S. (2020).
\newblock A new method to control error rates in automated species
  identification with deep learning algorithms.
\newblock {\em Sci. Rep.}, 10(1):10972.

\bibitem[Waide et~al., 2017]{waide2017demystifying}
Waide, R.~B., Brunt, J.~W., and Servilla, M.~S. (2017).
\newblock {Demystifying the landscape of ecological data repositories in the
  United States}.
\newblock {\em BioScience}, 67(12):1044--1051.

\bibitem[Weinstein, 2018]{Weinstein2018-kj}
Weinstein, B.~G. (2018).
\newblock Scene‐specific convolutional neural networks for video‐based
  biodiversity detection.
\newblock {\em Methods Ecol. Evol.}, 9(6):1435--1441.

\bibitem[Weinstein et~al., 2019]{Weinstein2019-wl}
Weinstein, B.~G., Marconi, S., Bohlman, S., Zare, A., and White, E. (2019).
\newblock Individual tree-crown detection in {RGB} imagery using
  semi-supervised deep learning neural networks.
\newblock {\em Remote Sens. (Basel)}, 11(11):1309.

\bibitem[Xie et~al., 2019]{Xie2019-gn}
Xie, J., Hu, K., Zhu, M., Yu, J., and Zhu, Q. (2019).
\newblock Investigation of different {CNN-based} models for improved bird sound
  classification.
\newblock {\em IEEE Access}, 7:175353--175361.

\bibitem[Younis et~al., 2020]{Younis2020-pz}
Younis, S., Schmidt, M., Weiland, C., Dressler, S., Seeger, B., and Hickler, T.
  (2020).
\newblock Detection and annotation of plant organs from digitised herbarium
  scans using deep learning.
\newblock {\em Biodivers. Data J.}, 8:e57090.

\bibitem[Zizka et~al., 2021]{Zizka2021-qc}
Zizka, A., Silvestro, D., Vitt, P., and Knight, T.~M. (2021).
\newblock Automated conservation assessment of the orchid family with deep
  learning.
\newblock {\em Conserv. Biol.}, 35(3):897--908.

\bibitem[Zualkernan et~al., 2020]{Zualkernan2020-hb}
Zualkernan, I.~A., Dhou, S., Judas, J., Sajun, A.~R., Gomez, B.~R., Hussain,
  L.~A., and Sakhnini, D. (2020).
\newblock Towards an {IoT-based} deep learning architecture for camera trap
  image classification.
\newblock In {\em 2020 {IEEE} Global Conference on Artificial Intelligence and
  Internet of Things ({GCAIoT})}. IEEE.

\end{thebibliography}

\end{document}